\documentclass[twoside,journal]{IEEEtran}

\usepackage{cite,graphicx,amssymb,amsmath,color,textcomp,tabularx}
\newtheorem{theorem}{\underline{Theorem}}
\newtheorem{lemma}{\underline{Lemma}}
\newtheorem{remark}{\underline{Remark}}

\newtheorem{proposition}{\underline{{Proposition}}}
\IEEEoverridecommandlockouts

\usepackage{graphicx}
\usepackage{epstopdf}
\usepackage{stfloats}
\usepackage{algorithm,algorithmic}

\usepackage{bm}
\usepackage{multirow}
\usepackage[caption=false,font=footnotesize]{subfig}

\usepackage{enumerate,extarrows}

\def \st {\mathrm{s.t.}}
\def \T {T} 

\def \E {\mathbb{E}}
\def \maxp {\mathop{\mathrm{maximize}}}

\newcommand{\mb}[1]{\mathbf{#1}}

\newcommand{\mr}[1]{\mathrm{#1}}

\usepackage{dsfont}

\newcommand{\HL}[1]{#1}

\begin{document}
\title{Wireless Powered Asynchronous Backscatter Networks with Sporadic Short Packets: Performance Analysis and Optimization}
\author{Qian~Yang,~\IEEEmembership{Student Member,~IEEE,}
	Hui-Ming~Wang,~\IEEEmembership{Senior Member,~IEEE,}\\
	Tong-Xing~Zheng,~\IEEEmembership{Member,~IEEE,}
	Zhu~Han,~\IEEEmembership{Fellow,~IEEE,}
	and~Moon Ho Lee,~\IEEEmembership{Life Senior Member,~IEEE}
	\thanks{Q. Yang, H.-M. Wang, and T.-X. Zheng are with the School of Electronic and Information Engineering, Xi'an Jiaotong University, Xi'an 710049, China, and also with the Ministry of Education Key Laboratory for Intelligent Networks and Network Security, Xi'an Jiaotong University, Xi'an 710049, China (e-mail: yangq36@gmail.com; xjbswhm@gmail.com; txzheng@stu.xjtu.edu.cn).}
	\thanks{Z. Han is with the University of Houston, Houston, TX 77004, USA (e-mail:zhan2@uh.edu), and also with the Department of Computer Science and Engineering, Kyung Hee University, Seoul, South Korea.}
	\thanks{M. H. Lee is with the Division of Electronics Engineering, Chonbuk National University, Jeonju 561-756, South Korea (e-mail: moonho@jbnu.ac.kr).}
	}

\maketitle

\begin{abstract}
	In the fifth generation era, the pervasive applications of Internet of Things and massive machine-type communications have initiated increasing research interests on the backscatter wireless powered communication (B-WPC) technique due to its ultra high energy efficiency and low cost. The ubiquitous B-WPC network is characterized by nodes with dynamic spatial positions and sporadic short packets, of which the performance has not been fully investigated. In this paper, we give a comprehensive analysis of a multi-antenna B-WPC network with sporadic short packets under a stochastic geometry framework. By exploiting a time-space Poisson point process model, the behavior of the network is well captured in a decentralized and asynchronous transmission way. We then analyze the energy and information outage performance in the energy harvest and backscatter modulation phases of the backscatter network, respectively.	The optimal transmission slot length and division are obtained by maximizing the network-wide spatial throughput. Moreover, we find an interesting result that there exists the optimal tradeoff between the durations of the energy harvest and backscatter modulation phases for spatial throughput maximization. Numerical results are demonstrated to verify our analytical findings and show that this tradeoff region gets shrunk when the outage constraints become more stringent.
\end{abstract}



\section{Introduction}
The promising Internet of Things (IoT) paradigm aims to bridge diverse technologies to support intelligent decision making by connecting pervasive physical objects together\cite{Al-Fuqaha2015}. 
One of the most important technologies for enabling IoT is backscatter radio, which has received increasing attention from both academic and industrial communities nowadays due to its distinguished low-energy requirement and rapid cut down of manufacture cost\cite{Xie2014}.
The most prominent commercial application of backscatter radio lies in radio frequency identification (RFID), which is the first technology realizing the machine-to-machine (M2M) concept (RFID tag and reader) and plays an indispensable role in the evolution of the IoT\cite{Al-Fuqaha2015}.
A typical RFID system consists of a RFID reader (interrogator) and a RFID tag (transponder) where the RFID reader interrogates the RFID tag for the desired information like an identification number\cite{Dobkin2007}.
Among all kinds of RFID tags, the \emph{passive} tag receives more emphasis compared with the active and semi-passive tags, since it has no on-tag power source but relies on the electromagnetic (EM) field transmitted by a RFID reader and ``backscattering'' for energy harvesting and information transmission, respectively\cite{Chawla2007}.
An illustration of passive tag applications is the wireless integrated sensing platform (WISP)\cite{Sample2007}, which is a programmable sensor platform based on RFID and aims to create a wirelessly-networked, battery-less sensor device. 
In the revolution of replacing the conventional barcode system, the commercial adoption of RFID is largely predicated on the passive tag due to its high energy efficiency, tiny size, and ultra low cost\cite{Boyer2014}.

In a passive RFID system, to initiate a querying procedure the RFID reader first transmits a continuous standardized signal to power up the passive RFID tag. Then the RFID tag wakes up after harvesting sufficient energy from the RFID reader's radio frequency (RF) signal and responds to the RFID reader via backscatter modulation. Finally, the RFID reader extracts the tag information from the signal backscattered by the passive RFID tag\cite{Dobkin2007}.
With the prevalence of the above described RFID backscatter system, there emerge the following three main research branches in the backscatter communication system.

The first and most important research branch is the transmit \emph{signal design and performance analysis} in the point-to-point backscatter system. To meet the increasing expectation of the data rate and data reliability for prominent RFID applications, many efforts have been devoted to investigating and improving the performance of the backscatter RFID system under multi-antenna frameworks \cite{Ingram2001,Griffin2008,Boyer2013,Boyer2014,He2014,He2015,He2016}.
The authors in \cite{Ingram2001} show that equipping with multiple antennas at both the RFID reader and tag can improve the coverage and capacity of the backscatter system.
The fading characteristics of the multiple-input multiple-output (MIMO) backscatter system is studied in \cite{Griffin2008} where the dyadic backscatter channel (DBC) is first introduced.
From then, the space-time coding scheme is widely explored in \cite{Boyer2013,Boyer2014,He2014,He2015,He2016} for MIMO RFID backscatter systems.
In addition, the transmit signal design of the RFID reader is investigated in \cite{Zheng2012}. With security issues further taken into account, the artificial-noise-aided transmit design is studied in \cite{Saad2014,Yang2016,Yang2016b} under the framework of physical layer security.
Recently, the ambient backscatter technology, which utilizes RF signals from the environment rather than the RF signals transmitted by the RFID reader, is studied in\cite{Kellogg2014,Wang2016a}.

Since the passive tag has no on-tag power source but relies on energy harvesting, the second research direction lies in considering \emph{energy transfer} in the point-to-point backscatter system, which is also called the backscatter wireless powered communication (B-WPC) system\cite{Huang2016}.
In \cite{Arnitz2013}, a measurement-based proof of the B-WPC concept is presented, and it is shown that the wireless energy transfer (WET) can be optimized by relying on only the power levels received by the base station.
The authors in \cite{Yang2015} study RF-enabled WET via energy beamforming in a backscatter system with multiple RFID tags, and a customized method is proposed to resolve the associated channel estimation problem.
However, the research on B-WPC is still sparse.

Different from the above research on \emph{point-to-point} backscatter communication systems, the third research trend arises from the increasing deployments of \emph{B-WPC networks}.
As it is pointed out in \cite{Xu2011}, handheld and mobile RFID readers become more and more prevailing due to their portability and flexibility in the evolution of the IoT in the fifth generation (5G) era.
Therefore, it is of great importance to further study the backscatter system design and B-WPC from the perspective of \emph{large-scale random networks}.
However, under pervasive and dense B-WPC networks in IoT and massive machine-type communications (mMTC)\cite{Bockelmann2016,Shariatmadari2015} applications such as in retail business, smart senors, and transportation checkpoints, how to formulate and resolve scalable and efficient WET and communication connectivity issues remains nearly untouched by the existing work.
Apart from the common features like the nodes with \emph{random spatial positions} in conventional large-scale networks, one of the most challenging characteristics brought by B-WPC networks in mMTC scenarios is that the transmissions are with \emph{sporadic short packets}\cite{Bockelmann2016}. 
The traffic of the nodes in these B-WPC networks is usually event-driven and uncoordinated, which makes slot-aligned transmissions prohibitive due to the overwhelming overhead for slot synchronization. Therefore, the grant-free access control is favorable for these networks and the packet transmission of the nodes in the whole network is usually asynchronous, which further incurs time-varying interference in the network. 
The study on how this time-varying interference affects the performance of the network-wide spatial throughput is challenging but interesting, since the interference is a contributing factor for WET while it hinders information reception.

\HL{%
\subsection{Related Work and Motivation}\label{subsec_related_work}
As reviewed in \cite{Liu2017b}, there exist lots of IoT applications for backscatter communications. For example, backscatter combined with WET provides a ultra low-cost and battery-free platform for the applications of smart homes or cities such as people or car navigation, weather monitoring, and security surveillance. Other application scenarios in IoT can be found in logistic tracking and biomedical applications \cite{Liu2017b}.
In \cite{Liu2017}, a multiple-access scheme based on time-hopping spread-spectrum is proposed to enable both one-way wireless energy transfer and two-way wireless information transfer under a system with coexisting backscatter reader-tag links.
However, the considered system model in \cite{Liu2017} is not general with only a limited number of predetermined backscatter reader-tag links, while the model with coexisting reader-tag pairs is quite common in practical IoT applications.
For instance, multiple customers in a shopping mall or supermarket may use their smart devices (i.e., RFID readers) to query prices or track producing information (contained in RFID tags) from various goods on the shelf at the same time, which actually constitutes a B-WPC interference network.

To cope with the feature of dynamic spatial positions of nodes in the random and large-scale network, the recently developed stochastic geometry theory has provided a powerful framework for evaluating the average performance of a dynamic network.
Under the stochastic geometry framework, the locations of the network nodes are modeled according to a certain spatial distribution such as the Poisson point process (PPP) and then the analytical analysis can be carried out easily\cite{Haenggi2009}.
Moreover, once the model parameters like network density are carefully chosen, these theoretical models can provide an accurate approximation for the network performance in real-world applications \cite{Andrews2011,ElSawy2017}.
We notice that the stochastic geometry model has been widely exploited to study the performance of WPC networks\cite{Bi2015,Huang2014,Huang2013,Che2015,Liu2016b} but not the backscatter applications.
In \cite{Huang2014}, the performance of microwave power transfer is investigated in an uplink cellular network powered by randomly deployed power beacons.
The network throughput maximization problem is studied under an energy harvesting mobile ad hoc network (MANET) and under a WPC network in \cite{Huang2013} and \cite{Che2015}, respectively.
The impact of physical layer security is further studied in \cite{Liu2016b}.
However, these work cannot be directly exploited in B-WPC applications with low-complexity passive devices due to the special mechanism of backscatter communications.

Despite the above-mentioned prominent applications in IoT, the area of the large-scale B-WPC network with multiple reader-tag links has been largely uncharted. A very recent contribution \cite{Han2017} considers the network coverage probability and transmission capacity of a B-WPC network with power beacons (PBs), where the network is modeled as a random Poisson cluster process by leveraging stochastic geometry.
Most recently, the network model in \cite{Han2017} is further extended to a more generic network setup with the energy from multiple nearby PBs taken into consideration in \cite{Bacha2017}.
However, the work in \cite{Han2017} and \cite{Bacha2017} only focuses on the scenario where the energy is harvested from dedicated power beacons, while the performance of the decentralized B-WPC MANET remains unknown. Most importantly, all the above work fails to capture the \emph{sporadic short-packet} nature of the B-WPC network in pervasive IoT and mMTC applications with asynchronous transmissions and the incurred time-varying interference.
The above reasons motivate our work.
}%

\subsection{Our Work and Contributions}\label{subsec_our_work}
In this paper, to better characterize stochastic networks with sporadic short packets, a time-space Poisson point process (TS-PPP) is exploited to model a B-WPC MANET. All the nodes in the network work in a decentralized and asynchronous way.
Different from \cite{Han2017}, we assume that each RFID reader is equipped with multiple antennas for more efficient power transfer and signal reception.
We give a comprehensive performance analysis of a B-WPC MANET and uncover the main tradeoff as well as the optimal parameter design in terms of spatial throughput maximization in the network.
The novelties and main contributions of this paper are summarized as follows:
\begin{enumerate}[1)]
	\item \HL{We study the performance of B-WPC in a multi-antenna MANET and introduce a TS-PPP model to fully capture the two main characteristics of the nodes in B-WPC networks in pervasive IoT and mMTC applications, i.e., dynamic spatial positions and sporadic short packets.}
	\item Based on the TS-PPP model, we analyze the energy and information outage performance arising in the two phases of each information transmission slot, namely the energy harvest and backscatter modulation phases, respectively. Moreover, the energy and information outage probabilities are derived in closed forms, which facilitates the further evaluation of network-wide metrics.
	\item We investigate the network-wide spatial throughput, which is defined as the total bits successfully transmitted by the RFID reader-tag pairs emerged at the unit time and network area. The spatial throughput is maximized by optimizing the slot length and slot division, i.e., the durations of the two phases, and we find an interesting result that there exists a tradeoff between the durations of the two phases for spatial throughput maximization. In addition, the optimal network time-space density is numerically analyzed for spatial throughput maximization when the slot length and slot division are predetermined.
\end{enumerate}

\subsection{Organization and Notations}
The remainder of this paper is organized as follows: In Sections \ref{sec_model} and \ref{sec_trans_model}, we present the network models and develop the formulation for the spatial throughput of the B-WPC MANET based on the information transmission model, respectively.
In Section \ref{sec_outage}, we investigate energy outage probability and information outage probability, respectively.
In Section \ref{sec_st_max}, we solve the spatial throughput maximization problem.
Numerical simulations and analysis are presented in Section \ref{sec_sim} before the conclusions drawn in Section \ref{sec_conclusion}.

\emph{Notations:} $\mathbf{A}^T$ and $\mathbf{A}^H$ represent the transpose and conjugate transpose of a matrix $\mathbf{A}$, respectively. $\mathbf{I}$ denotes an identity matrix.  $\mathbb{E}\{\cdot\}$ and $\mathbb{D}\{\cdot\}$ denote the expectation and variance operations, respectively.
$\mathds{1}(\cdot)$ denotes the indicator function.
$\mathbf{x}\sim \mathcal{CN}(\mathbf{\bm{\mu}},\mathbf{\Sigma})$ means that $\mathbf{x}$ is a random vector following a complex circular Gaussian distribution with mean $\mathbf{\bm{\mu}}$ and covariance $\mathbf{\Sigma}$.
$X\sim\mathrm{Exp}(\lambda)$ denotes the exponential distributed random variable with rate $\lambda$, and $X\sim\mathrm{Gamma}(k,\theta)$ denotes the Gamma-distributed random variable with shape $k$ and scale $\theta$. 
$\Gamma(x)$ is the Gamma function \cite[Eq. (8.310)]{Gradshteyn2007}, and $\gamma(k,x)$ is the lower incomplete gamma function \cite[Eq. (8.350.1)]{Gradshteyn2007}.

\section{Network Models}\label{sec_model}
\begin{figure}[t]
\centering
\includegraphics[width=3.5in]{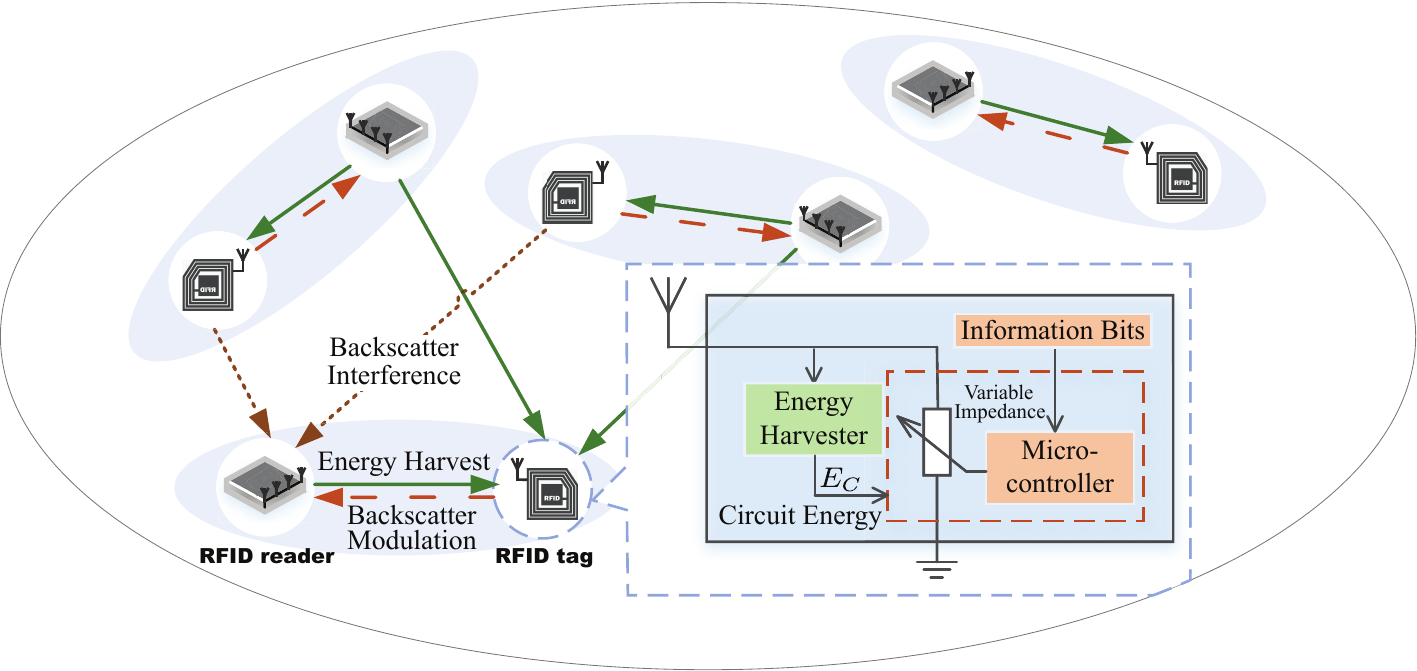}
\caption{The stochastic B-WPC ad hoc network consisting of multiple RFID reader-tag pairs.
	}
\label{fig_topo}
\end{figure}
We consider a stochastic B-WPC ad hoc network consisting of multiple RFID reader-tag pairs, where each reader and tag are separated by a common distance $d_0\geq 1$ as shown in Fig. \ref{fig_topo}.\footnote{Note that the results obtained in this paper can be extended to a general distance distribution in the similar way. The reason to adopt this simplified assumption is for the ease of analysis while preserving the key characteristics and tradeoffs in the network.}
As in \cite{Griffin2008,Boyer2013,Boyer2014}, each RFID reader in the network works in a full-duplex mode and is equipped with $M$ antennas for simultaneously transmitting query signal and receiving information signal from its corresponding RFID tag.
In addition, we consider a practical assumption that each RFID tag in the network is equipped with a single antenna.
Since \emph{passive} RFID tags are prominent for their low-cost and low-power characteristics\cite{Boyer2014}, we assume that all the RFID tags in the considered network are passive RFID tags.
In this paper, we focus on the information transmission from each RFID tag to its corresponding RFID reader, but before that each RFID tag has to be first powered up by harvesting the energy of the RF signals transmitted by the ambient RFID readers.
Therefore, each transmission slot for each RFID tag is comprised of the successive two phases, i.e., the energy harvest phase and the backscatter modulation phase.
For notational simplicity, we use the terms ``reader'' and ``tag'' as commonly used in RFID systems hereinafter.

\subsection{Asynchronous Network Model}\label{subsec_net_model}
In most of conventional network settings, it is assumed that the beginning time of each transmission slot for multiple transmitter-receiver pairs should be perfectly aligned. However, this assumption may be impractical especially for some decentralized networks like the ad hoc network where the overhead for the time slot synchronization is overwhelming and unaffordable.
Most importantly, with the coming of the 5G era pervasive IoT and mMTC applications in large-scale networks are featured with uncoordinated and sporadic short packets\cite{Bockelmann2016}, which makes the fulfillment of slot synchronization in these B-WPC networks even harder.
In this paper, we consider a more practical scenario where each reader-tag pair in the network randomly accesses the same frequency for ad hoc transmission in an asynchronous transmission way.

Different from the conventional stochastic geometry framework, the time randomness has to be further taken into consideration such that the asynchronous behavior of sporadic short packets in B-WPC networks can be better captured.
In this paper, we consider a homogeneous \emph{time-space} PPP $ \Phi=\{(x,t_x)\} $ of intensity $\lambda$ accounting for all the readers in the network, where $x\in \mathbb{R}^2$ denotes the location of a reader whose transmission starts from time $t_x$ and lasts for a predefined common time $T$.
In the considered network, it is assumed that there exists one tag paired with each reader according to some predetermined pairing rule\footnote{
	\HL{%
		To preserve the tractability of the main performance and tradeoff analysis in the considered network, as in \cite{Liu2017,Han2017,Bacha2017} the study of the specific association rule between readers and tags is out of the scope of this paper.
}%
}, and each tag is located at a distance $d_0$ from its paired reader in a random direction.
According to the displacement theorem\cite{Haenggi2012}, the collection of the corresponding tags is denoted by another TS-PPP $ \hat{\Phi}=\{(\hat{x},t_x)\} $ with intensity $\lambda$ where $\hat{x}\in \mathbb{R}^2$ is the location of the corresponding tag paired with the reader located at $x$.
This TS-PPP model seizes the essential time-space features of the network by leveraging only a single parameter $\lambda$, which captures the space-time frequency of the channel access, i.e., the number of transmission initiations per unit of space and time. Each reader-tag pair is born at a random time $t_x$ and random locations $x,\hat{x}$, accesses the channel during a time slot of duration $T$, and disappears immediately later. This model is also named as the Poisson rain model in \cite{Blaszczyszyn2010,Munari2015}.
Note that this model is naturally motivated by the two main characteristics of B-WPC networks in pervasive IoT and mMTC applications in the 5G era, i.e., dynamic spatial positions captured by $\{x,\hat{x}\}$ and sporadic short packets captured by $\{t_x\}$. Therefore, the model actually describes the behavior of the whole B-WPC network in a precise manner.

Under this stochastic geometry framework, the number of reader-tag pairs $N$ established communication in a region of area $A$ over time duration $T_0$ follows a Poisson distribution with parameter $\lambda A T_0$\HL{	\cite[Section 2.4.3]{Haenggi2012}}, namely
\begin{align}\label{Pr_N}
\Pr\{N=n\}=\frac{(\lambda A T_0)^n e^{-\lambda A T_0}}{n!}.
\end{align}

\subsection{Channel Model}\label{subsec_ch_model}
Based on the fact that the reader is usually close to its tag (small $d_0$) and beamforming is employed for efficient WPT, the forward channel from each reader to its corresponding tag is characterized by a path loss combined with a multi-antenna gain $G$ but no small-scale fading \cite{Huang2014,Huang2015,Han2017}, which is represented as $\sqrt{G}d_{0}^{-\alpha/2}$.
All the other channels, including interference channels from the concurrent transmission of the other reader-tag pairs and the reverse tag-reader channels of all the pairs, are assumed to undergo flat Rayleigh fading with a large-scale path loss governed by the exponent $\alpha>2$.\footnote{The analysis for different path-loss exponents accounting for WPT and backscatter information transmission can be performed in the similar way, which is omitted in this paper for the ease of presentation.}
To be specific, the forward channel from the reader located at $z$ to the tag located at $\hat{x}$ is represented as $\mb{h}_{z\hat{x}}d_{z\hat{x}}^{-\alpha/2}$, where $\mb{h}_{z\hat{x}}\in \mathbb{C}^{M\times 1}$ accounts for the small-scale fading with $\mb{h}_{z\hat{x}}\sim \mathcal{CN}(\mathbf{0},\mathbf{I})$ and $d_{z\hat{x}}$ denotes the path distance. Similarly, the reverse channel from the tag located at $\hat{x}$ to the reader located at $w$ is characterized by $\mb{g}_{\hat{x}w}r_{\hat{x}w}^{-\alpha/2}$ with $\mb{g}_{\hat{x}w}\sim \mathcal{CN}(\mathbf{0},\mathbf{I})$. 

\section{Information Transmission Model}\label{sec_trans_model}
As a passive tag, the fulfillment of passing the information such as identification data to its corresponding reader mainly relies on the procedure of \emph{backscatter modulation} which requires the passive tag to harvest sufficient energy first as illustrated in Fig. \ref{fig_topo}.

\begin{figure}[t]
	\centering
	\includegraphics[width=3.5in]{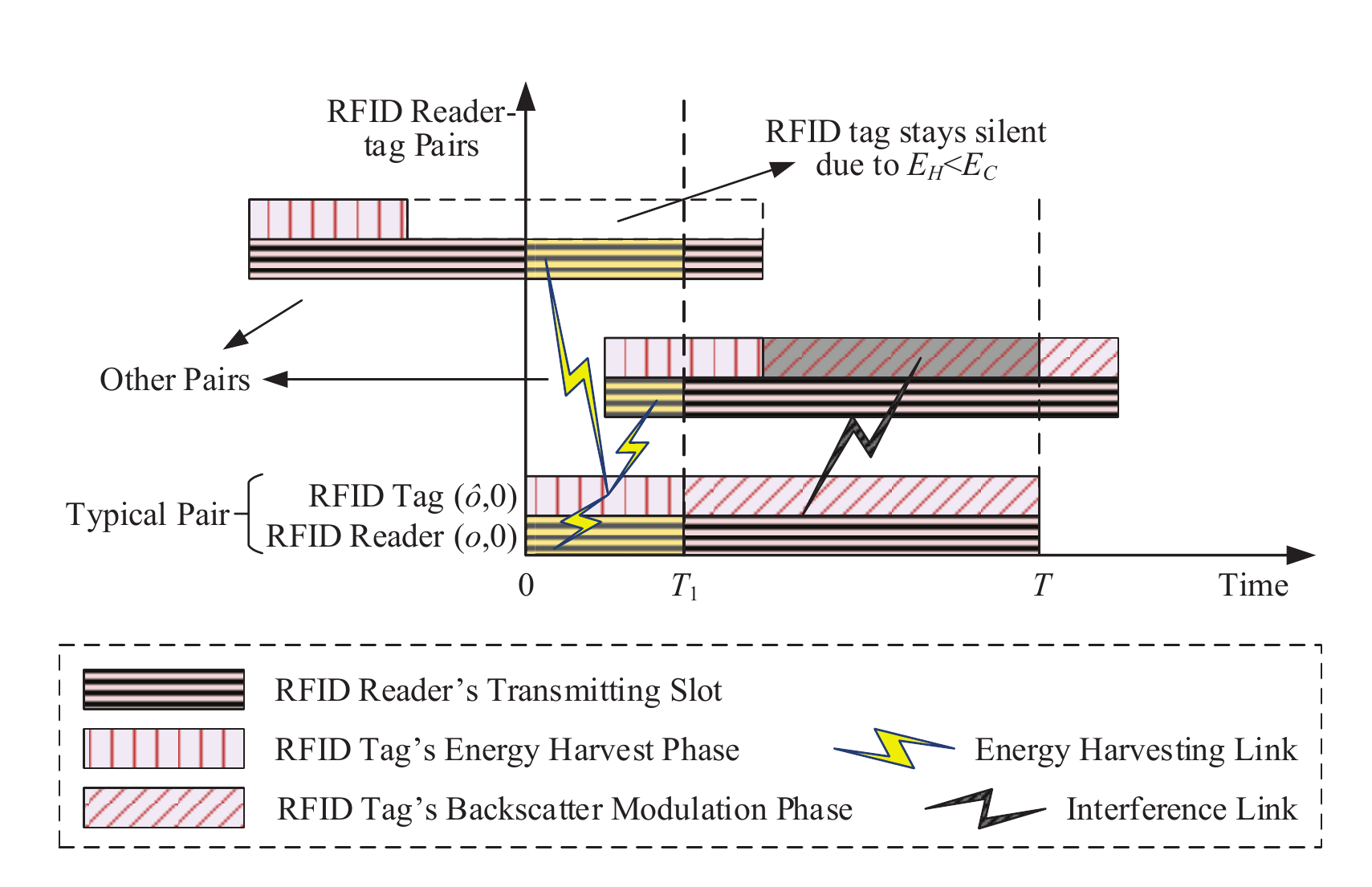}
	\caption{An illustration of the time slot model in the network.
	}
	\label{fig_slot}
\end{figure}

For the tractability of analysis, we consider a scenario where each tag in the network has a sporadic short packet with the same duration $T_2$ to transmit according to some unified protocol.
By splitting the duration of each transmission slot as $T=T_1+T_2$, each slot can be further divided into two phases, i.e., an \emph{energy harvest phase} of duration $T_1$ for energy harvesting and a \emph{backscatter modulation phase} of duration $T_2$ for packet transmission, respectively.
Note that different from conventional WET networks, in the considered B-WPC network the reader continuously transmits a standardized carrier-wave signal over the both phases in its slot. However, the signal acts as different roles during the two phases: an energy signal to power up the tag in the first phase and a query signal to enable the backscatter modulation in the second phase.
In the both phases, we assume that maximal-ratio transmit (MRT) is employed at each reader to improve the transmitting efficiency.
An illustration of the time slot model in the network is shown in Fig. \ref{fig_slot}.

\subsection{Energy Harvest Phase} \label{subsubsec_energy_phase}
By leveraging the homogeneity of the PPP\cite{Chiu2013}, 
we focus on a typical tag $(\hat{o},0)$ which locates at the origin and its reader begins the transmission at the time zero. 
During the energy harvest phase in time interval $[0, T_1]$, the tag harvests energy not only from its own dedicated reader $(o,0)$ but also from the other readers in the network which are concurrently transmitting.
Note that the transmitted signal from the other readers during either of their two transmission phases can be harvested to power up tag $(\hat{o},0)$ (cf. Fig. \ref{fig_slot}).
It is clear that the received power from the other readers at the typical tag is time-varying according to the assumed asynchronous network in Section \ref{subsec_net_model}.
For the ease of notation, we define $e(t,t_x)\triangleq \mathds{1}(t_x\leq t\leq t_x+T)$ with the aid of the indicator function.
\HL{%
	According to Fig. \ref{fig_slot}, the incident time-varying signal at the typical tag is given by 
	\begin{align} \label{harvest_power}
	y_T(t)=&\underbrace{\sqrt{ P_TGd_0^{-\alpha}} 
		e^{j\varphi_{\hat{o}}}}_{\text{from its paired reader}}\notag\\
	&+\underbrace{\sqrt{ P_T}\sum_{\substack{(x,t_x)\in\\ \Phi\backslash (o,0)}} e(t,t_x) \mb{h}_{x\hat{o}}^\T\mb{w}_{x} L(d_{x\hat{o}})^{-\alpha/2} e^{j\varphi_x} }_{\text{from other readers}}, 
	\end{align}
	where  %
	$P_T$ is the power of the energy signal transmitted by each reader,
	$\mb{w}_x=\mb{h}_{x\hat{x}}/\|\mb{h}_{x\hat{x}}\| $ is the MRT weight vector used at the transmitter of the reader $(x,t_x)$.
	Since the considered network is decentralized and the transmissions are asynchronous, the transmitted energy signals (i.e., the carrier-wave signals with amplitude $\sqrt{P_T}$) by different readers can have different initial phases.
	To model this, we use $\varphi_{\hat{o}}$ and $\varphi_x$ in \eqref{harvest_power} to account for different phases of the incident carrier-wave signal at tag $(\hat{o},0)$ from its paired reader $(o,0)$ and the transmitted carrier-wave signal by reader $(x,t_x)$, respectively, and these phases are assumed to be independently and uniformly distributed on $[-\pi,\pi)$.
	Note that in \eqref{harvest_power} we use a short-range propagation model for characterizing path loss with $L(d_{x\hat{o}})=\max\{d_{x\hat{o}},r_o\}$ and a constant $r_o\geq 1$ to avoid the singularity at zero distance and ensure that the finite moments of the total harvested energy exist as in \cite{Huang2014}.%
}%
\footnote{The tags in phase one also have the opportunity to harvest the backscattered power by the tags in phase two. However, this part of power is neglected in \eqref{harvest_power} based on the fact that the backscattered signal goes through a double path loss and is much smaller than the energy directly transferred from the readers.}
By leveraging \eqref{harvest_power}, the total harvested energy in the first phase is expressed as
\begin{align}\label{harvest_energy}
E_H=\eta \int_{0}^{T_1} |y_T(t)|^2 dt,
\end{align}
where $\eta$ is the energy-harvest efficiency\footnote{Note that a RF energy-harvesting sensitivity level is usually required to activate the harvesting circuit. However, we omit this level in this paper for the tractability of analysis as in \cite{Huang2014,Huang2013,Che2015,Liu2016b,Han2017} and for the fact that such sensitivity level can be kept relatively low compared to the harvested energy from the dedicated reader nowadays\cite{Lu2015}.
} at each tag.
If the harvested energy by the tag is larger than a given threshold $E_C$ at the end of the energy harvest phase, i.e., $E_H\geq E_C$, the tag will have sufficient energy to feedback the information back to its reader in the backscatter modulation phase. Otherwise, the tag will stay silent over the whole backscatter modulation phase as illustrated in Fig. \ref{fig_slot}\cite{Han2017}. 
The energy outage probability for a tag in the network is thereby given by
\begin{align}\label{P_eo}
P_{eo}=\Pr\{E_H<E_C\}.
\end{align}
Accordingly, the collections of the readers and the corresponding tags which have sufficient energy to transmit information at the second phase are denoted as $\Phi_T=\{(x,t_x)\in\Phi~|~E_H\geq E_C \}$ and $\hat{\Phi}_T=\{(\hat{x},t_x)\in\hat{\Phi}~|~E_H\geq E_C\}$ with the same intensity $\lambda_t\triangleq(1-P_{eo})\lambda$, respectively.

\begin{remark}
	It is worth noting that $\Phi_T$ and $\hat{\Phi}_T$ are no longer PPPs in general since the harvested energy at each tag is actually correlated by a common network of the readers and the thinning process is thereby dependent. However, it has been verified by the simulations in \cite{Che2015} that the thinned nodes still approximately form a PPP.
	For analytical tractability, we will view $\hat{\Phi}_T$ as a PPP in the sequel, and the accuracy of this approximation will be checked by the simulations in Section \ref{subsec_inf_out}.
\end{remark}

\subsection{Backscatter Modulation Phase} \label{subsubsec_backscatter_phase}
As illustrated in Fig. \ref{fig_topo}, when the passive tag has accumulated the sufficient energy $E_C$ at the beginning of the backscatter modulation phase, different impedance loads (or equivalently impedance mismatch levels) of its antenna are adjusted based on the stored information to reflect the incident query signal back to the reader. By doing so, the tag's information signal is modulated on the reflected query signal and finally is detected by the reader\cite{Boyer2014}.
\HL{%
	By similarly defining $ \iota(t,t_x)\triangleq \mathds{1}(t_x+T_1\leq t\leq t_x+T)$, according to Fig. \ref{fig_slot} the received signal at the typical reader $(o,0)$ which locates at the origin and starts the transmission at time zero is given by
	\begin{align} \label{y_R}
	&\mb{y}_R(t)=\underbrace{\left(y_T(t)\mb{g}_{\hat{o}o}d_0^{-\alpha/2}\right)\sqrt{\beta}s_{\hat{o}} \mathds{1}(T_1\leq t\leq T)}_\text{backscattered information signal from its paired tag}\notag\\
	&~~+ \underbrace{\sum_{\substack{(\hat{x},t_x)\in\\ \hat{\Phi}_T\backslash (\hat{o},0)}} \iota(t,t_x) \left(y_{T_{(\hat{x},t_x)}}(t) \cdot \mb{g}_{\hat{x}o}r_{\hat{x}o}^{-\alpha/2}\right)\sqrt{\beta}s_{\hat{x}}}_\text{interference from other tags} +\mb{n}, 
	\end{align}
	where $y_{T_{(\hat{x},t_x)}}(t)$ is the incident carrier-wave signal at tag $(\hat{x},t_x)$ and takes the similar form of \eqref{harvest_power} with $\hat{o}$ and $(o,0)$ changed to $\hat{x}$ and $(x,t_x)$, respectively;
	$s_{\hat{x}}$ denotes the information signal of tag $(\hat{x},t_x)$, which is modulated on the incident carrier wave by the tag and has unit power; $\beta$ is the backscatter coefficient of each tag during backscatter modulation,
	and $\mb{n}\sim \mathcal{CN}(\mathbf{0},N_0\mathbf{I})$ accounts for the additive white Gaussian noise (AWGN) at the receiver of each reader.
}%
It should be noted that the query signal terms transmitted from other readers do not appear in \eqref{y_R}, since the standardized  unmodulated carrier wave appears as a direct current (DC) level after down conversion and thereby can be easily removed\cite{Han2017}.
By denoting the signal-to-interference-plus-noise ratio (SINR) of the received signal in \eqref{y_R} as $\mr{SINR}_R$, the information outage probability for a reader in the backscatter modulation phase is thereby given by
\begin{align}\label{P_io}
P_{io}=\Pr\{\mr{SINR}_R<\gamma_R\},
\end{align}
where $\gamma_R$ is the targeted SINR level.

\subsection{Spatial Throughput Metric} \label{subsubsec_spatial_thr}
Similar to \cite{Huang2013,Che2015,Han2017}, we define the \emph{spatial throughput} of the backscatter network as the total bits successfully transmitted by the reader-tag pairs emerged at the unit time and network area.
Accordingly, the spatial throughput is characterized by
\begin{align} \label{R}
R(T_1,T_2,\beta)=\lambda_tT_2(1-P_{io})B,
\end{align}
where factor $B$ represents the data rate per tag-reader link during the backscatter modulation phase, and factor $\lambda_t (1-P_{io})$ accounts for the time-space density of active links with successful transmission. 
Note that the data rate $B$ is a function of the targeted SINR level $\gamma_R$, whose form is determined by the specific realization of modulation in practice.

\begin{remark}
	It is worth noting that to obtain the exact spatial throughput, the information outage probability $P_{io}$ used in \eqref{R} should be the probability conditioned on the reader-tag pairs which have harvested sufficient energy in the first phase. However, for analytical tractability we remark that imposing the independence on the two phases and simply substituting \eqref{P_io} into \eqref{R} actually incur little accuracy loss, which will be validated by the simulations in Section \ref{subsec_inf_out}.
\end{remark}

From \eqref{R}, we observe that the choices of $T_1$ and $T_2$ are critical for improving the spatial throughput.
Intuitively, larger $T_1$ or $T_2$ enables the tags to harvest more energy in the energy harvest phase, which improves the spatial throughput performance. However, larger $T_1$ or $T_2$ also incurs more interference to the readers in the backscatter modulation phase, which deteriorates the spatial throughput performance.
Therefore, it is of much interest to study how to design the transmission slot length and slot division to maximize the network-wide spatial throughput.
Under energy and information outage constraints, the spatial throughput maximization problem can be formulated as
\begin{equation}\label{max_st}
\begin{split} 
\maxp_{T_1,T_2}~~& R(T_1,T_2,\beta) \\
\st~~&P_{eo}\leq \epsilon_e, ~P_{io}\leq \epsilon_i.
\end{split}
\end{equation}
The above problem will be investigated in Section \ref{sec_st_max}.

\section{Energy and Information Outage Performance}\label{sec_outage}
In this section, we derive the analytical expressions for the energy and information outage probabilities during the energy harvest and backscatter modulation phases in Sections \ref{subsec_energy_out} and \ref{subsec_inf_out}, respectively.
\subsection{Energy Outage Probability} \label{subsec_energy_out}
Different from the scenario in the conventional WET network, calculating the energy outage probability given in \eqref{P_eo} is very challenging in the considered B-WPC network. 
The main reasons are summarized as follows:
\begin{enumerate}[1)]
	\item According to the transmission model given in \eqref{harvest_power}, the tags harvest energy from the carrier-wave signals with different phases transmitted by different readers in the network, which are actually highly correlated. Therefore, compared to the scenario in the conventional WET network, the total harvest power here is no longer simply the sum of different received signal power but the power of the received signal sum given by $|y_T(t)|^2$ as in \eqref{harvest_energy}. 
	\item Since the received power $|y_T(t)|^2$ is time-varying, the total harvested energy takes the integral form in \eqref{harvest_energy}, which is hard to tackle.
\end{enumerate}
The above two factors render finding the exact expression of the energy outage probability prohibitive.  

To circumvent this issue, we propose to exploit the \emph{second-order moment matching} technique to obtain an approximation of the probability. 
To be specific, the total harvested energy $E_H$ will be approximated as a Gamma random variable by matching the first and second order moments of $E_H$\cite{Heath2013}. Accordingly, the two parameters of the Gamma distribution is given in the following lemma.
\begin{lemma}\label{Lemma_gamma}
	After using second-order moment matching, the  distribution $\mathrm{Gamma}(k,\theta)$ with the same first and second order moments of $E_H$ has parameters
	\begin{align}
	k=\frac{\left(\mathbb{E}\{E_H\}\right)^2}{\mathbb{D}\{E_H\}},~\theta=\frac{\mathbb{D}\{E_H\}}{\mathbb{E}\{E_H\}}.
	\end{align}
\end{lemma}
\begin{IEEEproof}
	The lemma follows from \cite[Lemma 3]{Heath2013}.
\end{IEEEproof}

According to Lemma \ref{Lemma_gamma}, the mean and variance of $E_H$ have to be calculated in order to obtain an approximation of the energy outage probability. 
The results are provided in the following lemma.
\begin{lemma}\label{Lemma_EDX}
	The mean and variance of $E_H$ are given by
	\begin{align}\label{EX}
	\mathbb{E}\{E_H\}=\eta P_T T_1 \left(G d_0^{-\alpha}+\pi \lambda T \frac{\alpha}{\alpha-2} r_o^{2-\alpha}\right)
	\end{align}
	and 
	\begin{align}\label{DX}
	&\mathbb{D}\{E_H\}\notag\\
	&=(\eta P_T T_1)^2 \Bigg(\frac{2}{3} (2T_1+3T_2)\pi \lambda \alpha r_o^{2-\alpha} \left(\frac{Gd_0^{-\alpha}}{\alpha-2}+\frac{r_o^{-\alpha}}{\alpha-1}\right)  \notag\\
	&\quad+\left(\pi \lambda \frac{\alpha}{\alpha-2} r_o^{2-\alpha}\right)^2 \frac{1}{6} \left(3T_1^2+8T_1 T_2+6T_2^2\right)\Bigg),
	\end{align}
	respectively.
\end{lemma}
\begin{IEEEproof}
	The proof is given in Appendix \ref{app_Lemma_EDX}.
\end{IEEEproof}

It is worth noting that the first term in \eqref{EX} is actually the energy harvested directly from each tag's dedicated reader, while the second term in \eqref{EX} accounts for the average energy harvested from the B-WPC network. 
By leveraging Lemmas \ref{Lemma_gamma} and \ref{Lemma_EDX}, an approximation of the energy outage probability defined in \eqref{P_eo} is provided in the following theorem.
\begin{theorem}\label{th_energy_out}
	The energy outage probability for the tags in the energy harvest phase can be approximated as
	\begin{align}\label{P_eo_app}
	P_{eo}\approx \frac{\gamma(k,E_C/\theta)}{\Gamma(k)},
	\end{align}
	where $k$ and $\theta$ are given in Lemma \ref{Lemma_gamma}.
\end{theorem}
\begin{IEEEproof}
	By leveraging Lemma \ref{Lemma_gamma} and the cumulative distribution function (CDF) of Gamma random variables, the theorem is immediately obtained.
\end{IEEEproof}

In addition, lower and upper bounds of $P_{eo}$ can be obtained by using the Cantelli inequality (also known as the one-sided Chebyshev inequality\cite{Marshall1960}) and the results in Lemma \ref{Lemma_EDX}. To be specific, the energy outage probability is upper bounded as
\begin{align}\label{energy_ub}
P_{eo} \leq \frac{\mathbb{D}\{E_H\}}{\mathbb{D}\{E_H\}+(E_C-\mathbb{E}\{E_H\})^2}, ~~ \text{for}~ E_C\leq  \mathbb{E}\{E_H\},
\end{align}
and a lower bound of $P_{eo}$ is given by
\begin{align}\label{energy_lb}
P_{eo} \geq \frac{(E_C-\mathbb{E}\{E_H\})^2}{\mathbb{D}\{E_H\}+(E_C-\mathbb{E}\{E_H\})^2}, ~~ \text{for}~ E_C\geq  \mathbb{E}\{E_H\}.
\end{align}

\begin{figure}[t]
	\centering
	\includegraphics[width=3.5in]{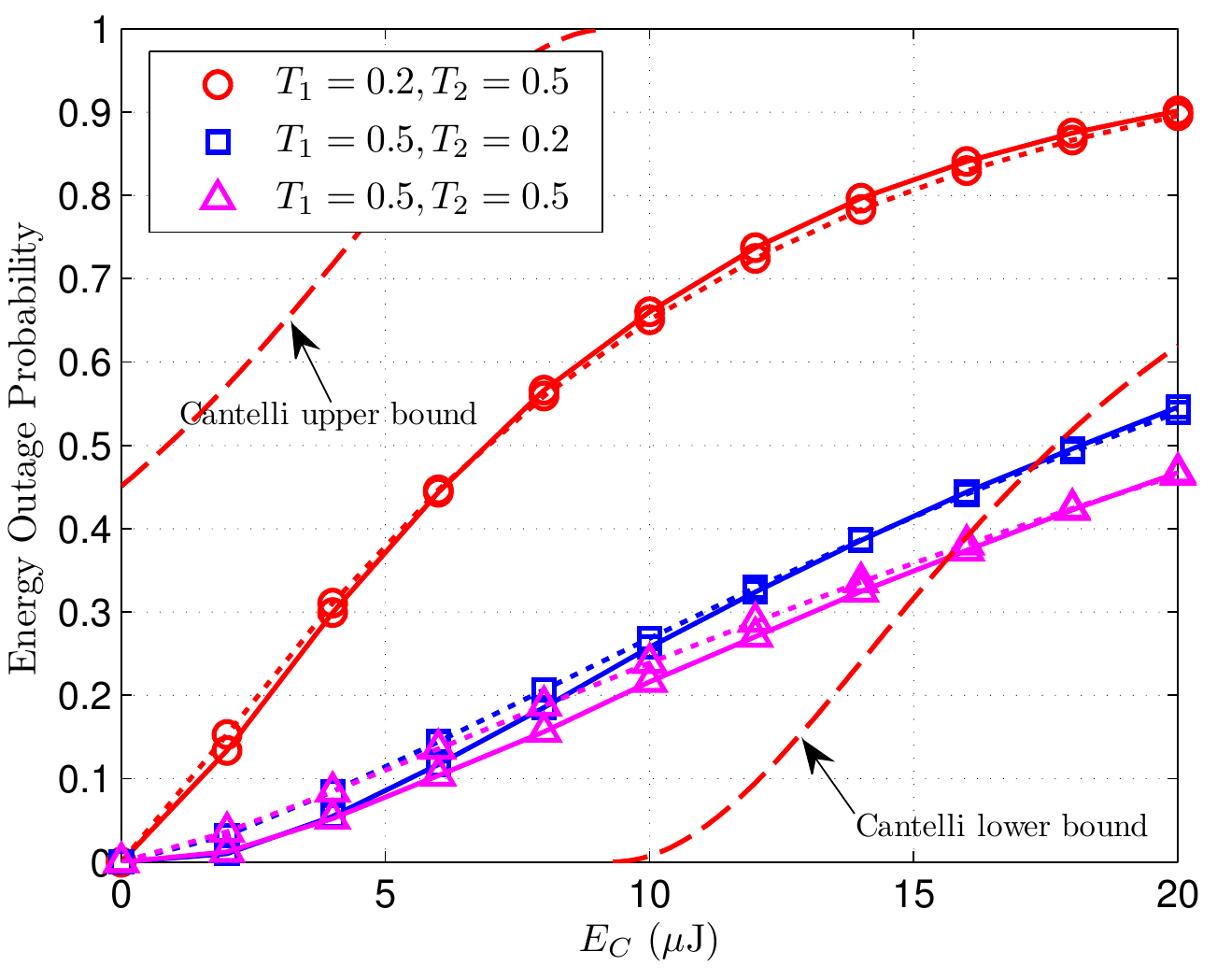}
	\caption{The energy outage probability versus the energy threshold $E_C$ in the energy harvest phase. The solid lines stand for simulation results, while dashed lines present the corresponding analytical results. Moreover, the upper and lower Cantelli bounds are provided for the case of $T_1=0.2~\mr{ms}$ and $T_2=0.5~\mr{ms}$. The other network parameters are: $\lambda=0.1~/(\mr{m}^2\cdot \mr{ms})$, $M=3$, $G=\varrho M$ with $\varrho=2/3$, $P_T=20$ dBm (100 mW), $\eta=0.8$, $d_0=r_o=2$ m, and $\alpha=3$.}
	\label{fig_Energy_Ec}
\end{figure}

\HL{%
	To validate our theoretical findings, we leverage the Monte Carlo method by averaging $10^3$ network realizations to obtain the simulation results throughout the paper. Note that finite yet adequately large network area and time duration are considered in the simulations such that the introduced border effect on the performance can be neglected.
}%
Fig. \ref{fig_Energy_Ec} depicts the energy outage probability versus the energy threshold $E_C$ under different network parameters, where the upper and lower Cantelli bounds are provided for the case of $T_1=0.2~\mr{ms}$ and $T_2=0.5~\mr{ms}$.
From Fig. \ref{fig_Energy_Ec}, we observe that the bounds given by \eqref{energy_ub} and \eqref{energy_lb} are loose, since the Cantelli inequality is general and holds for any random variables. 
Contrarily, our analytical approximations obtained by Theorem \ref{th_energy_out} are found very close to the simulation results. 
It can be seen from Fig. \ref{fig_Energy_Ec} that reducing $T_1$ or $T_2$ both leads to an increase in the energy outage probability. This is because when the whole transmission duration $T$ gets smaller, the tags in the first phase are likely to harvest less power from the ambient readers' signal which lasts for both the durations of $T_1$ and $T_2$. Above all, since $T_1$ directly determines the duration of the energy harvest phase, reducing $T_1$ impacts the energy outage performance even more.

\subsection{Information Outage Probability} \label{subsec_inf_out}
When the tag harvests sufficient energy in the first phase, i.e., there is no energy outage, the passive tag will pass its stored information to the reader during the backscatter modulation phase.
To calculate the information outage probability during this phase, we first formulate the SINR of the received signal at the typical reader in \eqref{y_R}.
For analysis tractability, we use the expectation of average power
\begin{align}
\bar{y}_T^2\triangleq \mathbb{E}\left\{\frac{1}{T_2}\int_{T_1}^{T} |y_T(t)|^2 dt\right\}= \frac{1}{\eta T_1}\mathbb{E}\{E_H\}
\end{align}
to approximate the power of $y_T(t)$ and $y_{T_{(\hat{x},t_x)}}(t)$ for any $(\hat{x},t_x)\in \hat{\Phi}_T\backslash (\hat{o},0)$ during time duration $[T_1,T]$. 
Note that this approximation is quite accurate when the incident power transferred from the dedicated reader is dominant in the total incident power at the tag, which is the usual case.
Let 
\begin{align}
\mb{R}_I(t)=\beta \bar{y}_T^2 \sum_{\substack{(\hat{x},t_x)\in \hat{\Phi}_T\backslash (\hat{o},0)}} \iota(t,t_x) r_{\hat{x}o}^{-\alpha} \mb{g}_{\hat{x}o}\mb{g}_{\hat{x}o}^H
\end{align}
be the covariance matrix of the received interference.
To further evaluate the information outage performance of the network, the time-varying term $\mb{R}_I(t)$ renders the $\mr{SINR}_R$ intractable. To circumvent this problem, as in \cite{Blaszczyszyn2010,Munari2015} we use the average interference value over the backscatter modulation phase $\bar{\mb{R}}_I=1/T_2\int_{T_1}^{T}\mb{R}_I(t)dt$ to replace the intractable time-varying term.
\HL{%
Note that this corresponds to the case where some coding with repetition and interleaving of bits is employed \cite{Blaszczyszyn2010}.
We only consider this average interference case in our paper, and other more sophisticated scenarios like the maximal interference constraint for the outage probability are left for future work.
}%
By employing the minimum mean square error (MMSE) receiver, the optimal SINR performance at the typical reader can be characterized as
\begin{align} \label{SINR_R}
&\mr{SINR}_R\notag\\
&=\beta \bar{y}_T^2 d_0^{-\alpha} \mb{g}_{\hat{o}o}^H(\bar{\mb{R}}_I+N_0\mb{I})^{-1} \mb{g}_{\hat{o}o} \notag\\
&=d_0^{-\alpha} \mb{g}_{\hat{o}o}^H\left(\sum_{\substack{(\hat{x},t_x)\in\\ \hat{\Phi}_T\backslash (\hat{o},0)}} w(t_x) r_{\hat{x}o}^{-\alpha} \mb{g}_{\hat{x}o}\mb{g}_{\hat{x}o}^H +\frac{N_0}{\beta \bar{y}_T^2 } \mb{I}\right)^{-1} \mb{g}_{\hat{o}o},
\end{align}
where
\begin{align} \label{w_tx}
w(t_x)\triangleq \frac{1}{T_2} \int_{T_1}^{T}\iota(t,t_x)dt =\begin{cases}
\frac{T_2-|t_x|}{T_2}, &t_x\in[-T_2,T_2],\\
0,&\mr{elsewhere}.
\end{cases}
\end{align}

To calculate the information outage probability defined in \eqref{P_io}, the main obstacles lie in the fact that the number of interference terms in \eqref{SINR_R} is infinite and the time randomness of the network $\{t_x\}$ involved in \eqref{w_tx} should be further tackled, which makes the calculation of \eqref{P_io} challenging.
\HL{%
To circumvent these problems, our method is to first calculate the information outage probability for a fixed area and time duration, and then we obtain the final result by letting the area and time duration go to infinity.
Note that this approach leads to no loss of accuracy.
}%
The main result is given in the following theorem.
\begin{theorem} \label{th_P_io}
	The information outage probability defined in \eqref{P_io} in the backscatter modulation phase is given by
	\begin{align}\label{P_io_res}
	P_{io}=1-H\left(\lambda_t(\gamma_R d_0^\alpha)^\delta\Delta T_2+\frac{N_0}{\beta \bar{y}_T^2 } \gamma_R d_0^\alpha\right),
	\end{align}
	where $H(x)\triangleq  \sum_{i=0}^{M-1} \frac{x^i}{i!} e^{-x}$, $\Delta\triangleq \frac{4\pi}{2+\alpha}\Gamma(\delta)\Gamma(1-\delta)=\frac{4\pi^2}{(2+\alpha)\sin(\delta\pi)}$, and $\delta\triangleq \frac{2}{\alpha}$.
\end{theorem}
\begin{IEEEproof}
	The proof is given in Appendix \ref{app_P_io}.
\end{IEEEproof}

Since the first order derivative of $H(x)$ can be found as
\begin{align}\label{H'_x}
H'(x)=-\frac{x^{M-1}}{(M-1)!}e^{-x}\leq 0,
\end{align}
the information outage probability decreases with an increase in the backscatter coefficient $\beta$ or the incident power $\bar{y}_T^2$.
This result is not surprising, because increasing $\beta$ or $\bar{y}_T^2$ leads to an increase in both the signal and interference terms in \eqref{y_R}, which is equivalent to reducing the AWGN level $\sigma^2\triangleq \frac{N_0}{\beta \bar{y}_T^2 }$ with the backscatter power normalized to one.

In realistic backscatter applications, the receiver noise power $N_0$ is small (as low as $-90$ dBm\cite{Saad2014}) and can be much lower than the power of network interference. Therefore, the study of the network performance under the interference-limited regime is of much importance. Under this scenario, the information outage probability in \eqref{P_io_res} becomes
\begin{align}\label{P_io_int}
P_{io}=1-H\left(\lambda_t(\gamma_R d_0^\alpha)^\delta\Delta T_2\right).
\end{align}
From \eqref{P_io_int}, based on \eqref{H'_x} 
and by recalling $\lambda_t=(1-P_{eo})\lambda$ where $P_{eo}$ decreases with an increase in $T_1$ or $T_2$,
we know that the information outage probability increases with $T_1$ and $T_2$. This is because the average number of emerged interfering nodes in unit area during the backscatter modulation phase becomes larger when $\lambda_tT_2$ increases.

\begin{figure}[t]
	\centering
	\includegraphics[width=3.5in]{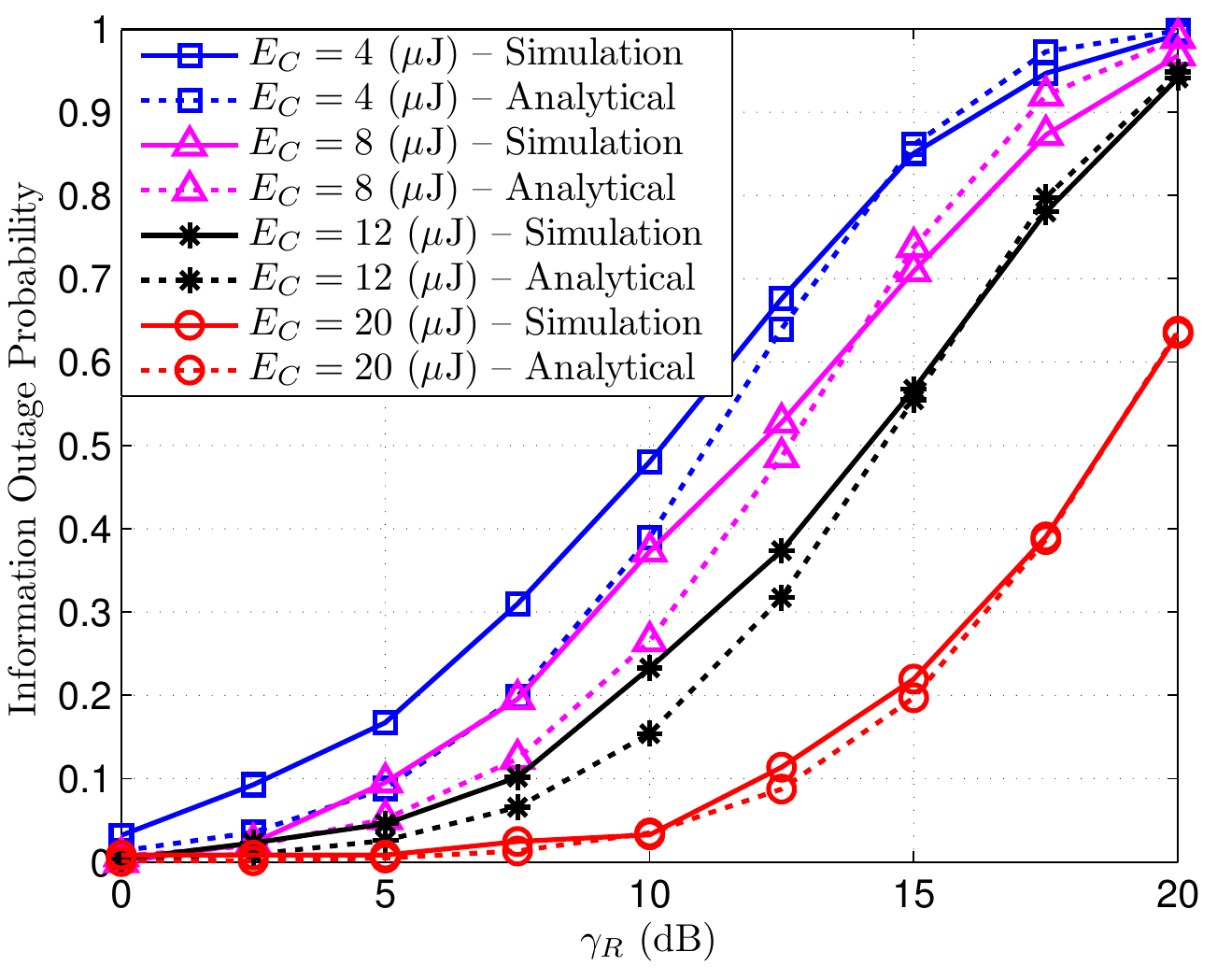}
	\caption{\HL{The information outage probability versus the targeted SINR level $\gamma_R$ in the backscatter modulation phase under different $E_C$'s. The other network parameters are: $\lambda=0.03~/(\mr{m}^2\cdot \mr{ms})$, $T_1=T_2=0.5$ ms, $N_0=-90$ dBm, $\beta=0.8$, $M=3$, $G=\varrho M$ with $\varrho=2/3$, $P_T=20$ dBm (100 mW), $\eta=0.8$, $d_0=r_o=2$ m, and $\alpha=3$.}}
	\label{fig_Inf_gamma}
\end{figure}
Fig. \ref{fig_Inf_gamma} depicts the relation between the information outage performance and the targeted SINR level under different energy thresholds. The simulation result shown in Fig. \ref{fig_Inf_gamma} is the information outage probability conditioned on the successful energy harvesting ($E_H\geq E_C$) in the first phase. The results are obtained by counting the information outage events of the tag located at the origin only when it has harvested sufficient energy in the first phase. As we can see from Fig. \ref{fig_Inf_gamma}, our analytical results coincide with the simulation ones, which validates the correctness of our analytical derivations.
In addition, it is straightforward to see that $P_{io}$ becomes smaller with a larger demanded $E_C$, since the working tags in the backscatter modulation phase become sparser under this situation.

\section{Spatial Throughput Maximization}\label{sec_st_max}
As pointed out in Section \ref{sec_outage}, the choices of $T_1$ and $T_2$ largely determine the performance of the energy and information outage in the two transmission phases.
Specifically, setting a larger $T_1$ or $T_2$ will be beneficial for energy harvest while it will deteriorate the performance of information decoding during the backscatter modulation phase.
Therefore, it is of great importance to investigate the optimal design for spatial throughput maximization and the optimal tradeoff between $T_1$ and $T_2$.

By focusing on the interference-limited regime and substituting \eqref{P_io_int} into \eqref{R}, we obtain the spatial throughput of the asynchronous B-WPC network as
\begin{align} \label{R1}
&R(T_1,T_2,\beta)=\lambda_tT_2 \cdot H\left(\lambda_tT_2 (\gamma_R d_0^\alpha)^\delta\Delta\right) \cdot B.
\end{align}
Based on \eqref{R1}, the spatial throughput maximization problem in \eqref{max_st} is equivalent to
\begin{subequations}\label{max_st1}
\begin{align} 
\maxp_{T_1,T_2,\chi}~~& f(\chi)\triangleq \chi \cdot H\left(\chi  \right) \label{obj}\\
\st~~&\chi= \tau \lambda_tT_2,\\
&P_{eo}\leq \epsilon_e, ~P_{io}\leq \epsilon_i, \label{out_constr}
\end{align}
\end{subequations}
where $\lambda_t$ is a function of both $T_1$ and $T_2$ by recalling that $\lambda_t=(1-P_{eo})\lambda$, and $ \tau \triangleq (\gamma_R d_0^\alpha)^\delta\Delta$.
Before proceeding, we first show that the objective function $f(\chi)$ in \eqref{obj} is quasi-concave\cite[p. 99]{Boyd2004} for $\chi\geq 0$ in the following proposition.
\begin{proposition}\label{prop_fx}
	The function $f(x)=xH(x)= \sum_{i=0}^{M-1} \frac{x^{i+1}}{i!} e^{-x}$ is quasi-concave for $x\geq 0$. Specifically, $f(x)$ first increases in $x\in[0,x_M]$ and then decreases in $x\in(x_M,\infty)$, where $x_M\in(\frac{M}{2},M)$ is the unique number satisfying
	\begin{align}\label{x_M} Q(x)=-\frac{x^M}{(M-1)!}+\sum_{n=0}^{M-1}\frac{x^n}{n!}=0.
	\end{align}
\end{proposition}
\begin{IEEEproof}
	Please refer to \cite[Appendix B]{Ali2010} for the detail of the proof. The quasi-concavity of $f$ is immediately obtained according to the definition of the quasi-concave function in \cite[p. 99]{Boyd2004}.
\end{IEEEproof}

It is worth noting that based on the quasi-concavity of function $f(x)$ in Proposition \ref{prop_fx}, the maximizer $x_M$ of $f$ can be efficiently calculated via a bisection search in the interval $(\frac{M}{2},M)$. 
Additionally, when the outage constraints in \eqref{out_constr} are absent, the optimal spatial density $\lambda_tT_2$ linearly scales with the antenna number at each reader $M$ due to $x_M\in(\frac{M}{2},M)$.

By leveraging the monotonic property of the information outage probability based on \eqref{H'_x}, problem \eqref{max_st1} is equivalent to
\begin{subequations}\label{max_st2}
\begin{align}
\maxp_{T_1,T_2,\chi}~~& f(\chi) \\
\st~~&\chi= \tau\lambda T_2 (1-P_{eo}), \label{cc_eq}\\
&\chi \leq \chi_\mr{up}, \label{cc_chi}\\
&P_{eo}\leq \epsilon_e, \label{cc_Peo}
\end{align}
\end{subequations}
where $\chi_\mr{up}\triangleq H^{-1}(1-\epsilon_i)$ with $H^{-1}(x)$ being the inverse function of $H(x)$.
By considering the constraint on $\chi$ in \eqref{cc_chi} and leveraging the quasi-concave property of $f(\chi)$ in Proposition \ref{prop_fx}, we obtain the optimal
\begin{align}
\chi^\star=\min\{x_M,\chi_\mr{up}\}
\end{align}
for problem \eqref{max_st2}.
By further jointly taking account of the remaining constraints in \eqref{cc_eq} and \eqref{cc_Peo}, the constraint on $P_{eo}$ in \eqref{cc_Peo} can be recast as $T_2\leq \frac{\chi^\star}{ \tau\lambda(1-\epsilon_e)}$. 
Therefore, the optimal solution $(T_1^\star,T_2^\star)$ to the spatial throughput maximization problem is given by the collection
\begin{align}\label{sol} 
&(T_1^\star,T_2^\star) \notag\\
&\in \left\{(T_1,T_2) ~\bigg|~T_2 \left(1-P_{eo}\right)=\frac{\chi^\star}{ \tau\lambda},~T_2\leq \frac{\chi^\star}{ \tau\lambda(1-\epsilon_e)} \right\},
\end{align}
and the optimal spatial throughput is thereby given by
\begin{align}\label{opt_thr}
R_\mr{opt}=\frac{\chi^\star}{\tau} \cdot H(\chi^\star) \cdot B.
\end{align}
Concerning the optimal solution $(T_1^\star,T_2^\star)$ given in \eqref{sol}, we have the following observations:

\subsubsection{Existence of the Optimal Solution}
The optimal solution collection in \eqref{sol} always exists. This observation can be easily verified as follows. From the equality constraint in \eqref{sol}, we obtain the lower bound of $T_2$ given by $T_2^\mr{LB}=\frac{\chi^\star}{ \tau\lambda}$ due to $P_{eo}\geq0$. By noting that this lower bound of $T_2$ is always smaller than its upper bound $T_2^\mr{UB}= \frac{\chi^\star}{ \tau\lambda(1-\epsilon_e)}$, the optimal $T_2^\star$ can be arbitrarily chosen from $(T_2^\mr{LB},T_2^\mr{UB}]$. Once $T_2^\star$ is determined, the optimal $T_1^\star$ can be obtained according to the equality constraint in \eqref{sol}.

\subsubsection{Tradeoff of the Optimal Solution}
From \eqref{sol}, we know that there exists a tradeoff between $T_1$ and $T_2$, while different choices of $T_1$ and $T_2$ restrained by \eqref{sol} yield the same optimal objective value.
To see more clearly, we rewrite \eqref{sol} as
\begin{align}\label{sol1}
\left\{(T_1,T_2) ~\bigg|~p(T_1,T_2) =\frac{\chi^\star}{ \tau\lambda},~T_2^\mr{LB}<T_2\leq T_2^\mr{UB} \right\},
\end{align}
where 
\begin{align}
p(T_1,T_2)\triangleq T_2 \left(1-P_{eo}(T_1,T_2)\right).
\end{align}
In light of the numerical results shown in Section \ref{subsec_energy_out}, it is not hard to see that the function $p(T_1,T_2)$ is an increasing function of both $T_1$ and $T_2$. 
This observation means that to achieve the same optimal spatial throughput, $T_1$ and $T_2$ can complement for each other.
For example, on the line constrained by $p(T_1,T_2) =\frac{\chi^\star}{ \tau\lambda}$, we can find a smaller $T_2$ by enlarging $T_1$ to achieve a tighter constraint for energy outage.
In addition, we remark that the obtained solution given in \eqref{sol} can further easily incorporate individual constraints on the time lengths of $T_1$ and $T_2$.

\subsubsection{Optimal Network Density for Spatial Throughput Maximization}
By far, we have found the optimal slot design $(T_1^\star,T_2^\star)$ which maximizes the spatial throughput. We remark that when the time durations of the two transmission phases are predetermined according to some given protocol, the similar tradeoff also exists in network density optimization for spatial throughput maximization. However, as will be observed by the simulations in Section \ref{sec_sim}, the energy outage probability $P_{eo}(\lambda)$ is not necessarily an decreasing function of $\lambda$ in the considered B-WPC network. Therefore, the similar method developed in this section cannot be used, and we will numerically check the impact of network density on the spatial throughput performance in Section \ref{sec_sim}.

\section{Numerical Results}\label{sec_sim}
In this section, more representative results are provided to unveil the network performance and optimal parameter design in terms of spatial throughput maximization.

\HL{%
	As in \cite{Han2017,Bacha2017}, the Monte Carlo simulations are conducted with the aid of the MATLAB software on a desktop computer.
}%
The parameter settings are as follows, unless otherwise specified.
The density of the network is $\lambda=0.03~/(\mr{m}^2\cdot \mr{ms})$.
The time durations allocated for the energy harvest and backscatter modulation phases in each slot are $T_1=T_2=0.5$ ms.
The antenna number at each reader in the network is $M=3$, and the multi-antenna gain for direct energy transfer is set as $G=\varrho M$ with $\varrho=2/3$.
The transmit power for each reader is $P_T=20$ dBm (100 mW), and the energy-harvest efficiency at each tag is $\eta=0.8$.
The distance between the reader and the tag in each pair is $d_0=2$ m, $r_o=2$ m, and the path-loss exponent is $\alpha=3$.
The relation between the data rate $B$ and the targeted SINR level $\gamma_R$ is captured by $B=\log(1+\gamma_R)$.


\begin{figure}[t]
	\centering
	\includegraphics[width=3.5in]{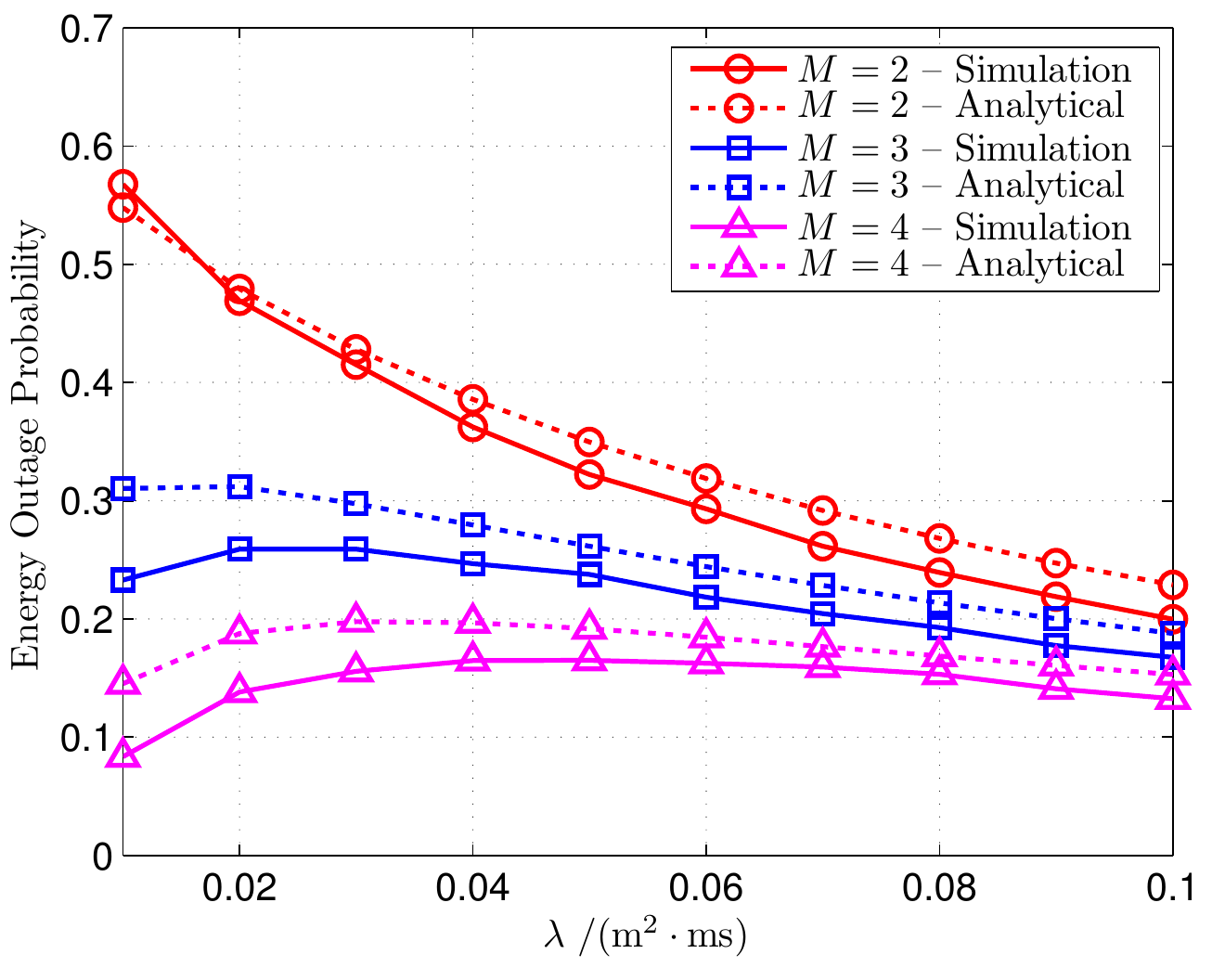}
	\caption{The energy outage probability versus the network density $\lambda$ in the energy harvest phase under different antenna numbers at the readers. The energy threshold is $E_C=8~\mu\mathrm{J}$.}
	\label{fig_Energy_l}
\end{figure}
We first check the impact of the network density $\lambda$ on the energy outage performance. 
As shown in Fig. \ref{fig_Energy_l}, the energy outage probability is not always a decreasing function of $\lambda$, e. g. the case of $M=4$. Note that this result is different from the one in the conventional WPC network where the WPC nodes always benefit more from denser network deployments. The reason behind is that the transmitted carrier-wave signal by different readers is actually highly correlated due to the same working frequency. 
Therefore, the harvested total energy is no longer simply the summation of the energy independently harvested from each network node. 
As observed from Fig. \ref{fig_Energy_l}, the energy outage probability first increases as $\lambda$ increases from 0.01 to 0.03 $/(\mr{m}^2\cdot \mr{ms})$ under a large antenna number $M=4$. This is because the outage probability is small due to the large multi-antenna gain provided by the dedicated reader, when $\lambda$ is extremely small and the energy harvested from ambient readers in the network can be neglected.
When $\lambda$ gets larger, the incident carrier waves from different readers can be destructively combined, which leads to a higher energy outage probability.
Furthermore, from Fig. \ref{fig_Energy_l} the outage probability decreases when the readers in the network are equipped with more antennas as expected.

\begin{figure}[t]
	\centering
	\subfloat[Mesh figure]{ 
		\includegraphics[width=3.5in]{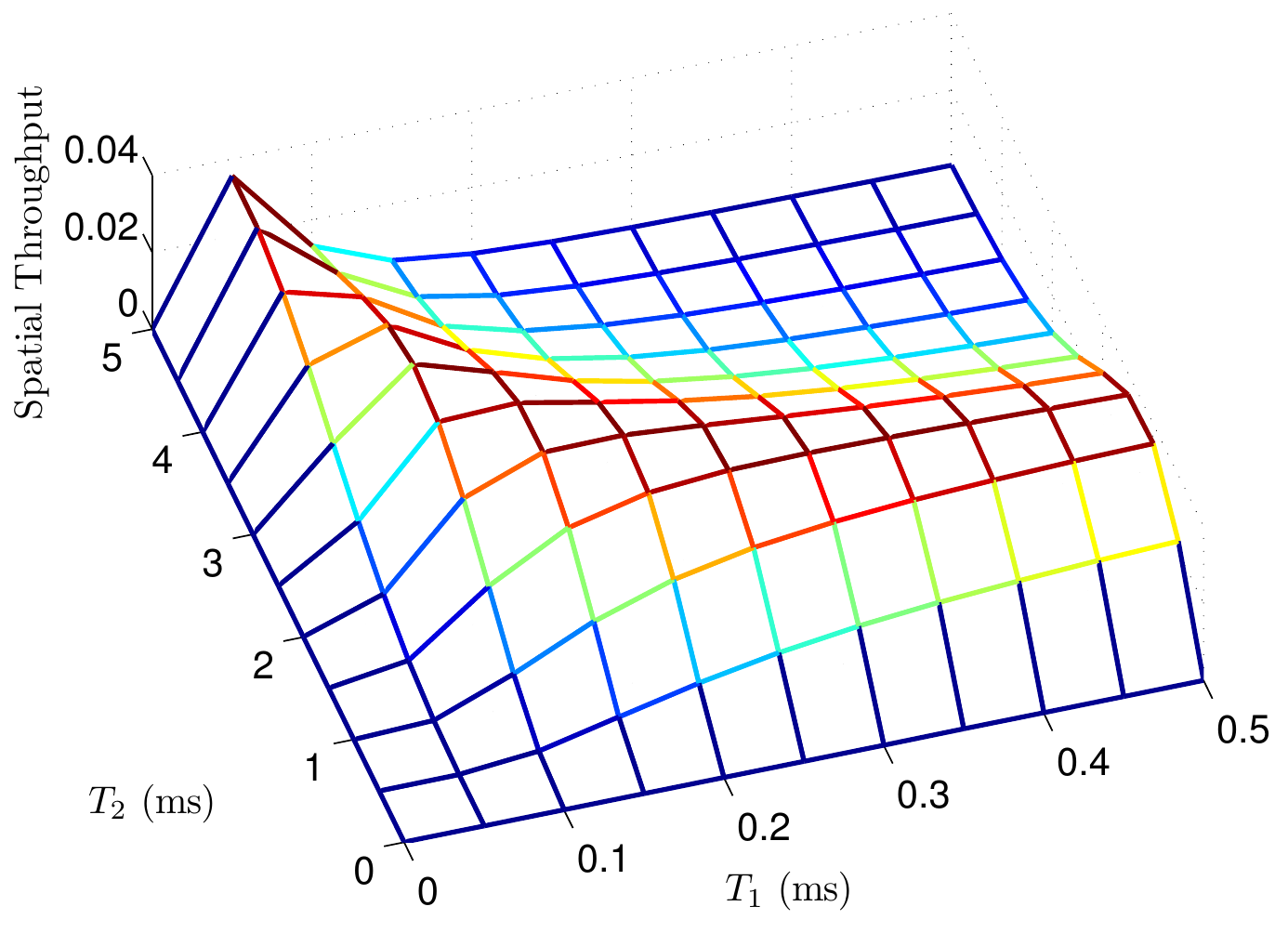}
	}
	\hfil
	\subfloat[Contour figure]{ 
		\includegraphics[width=3.5in]{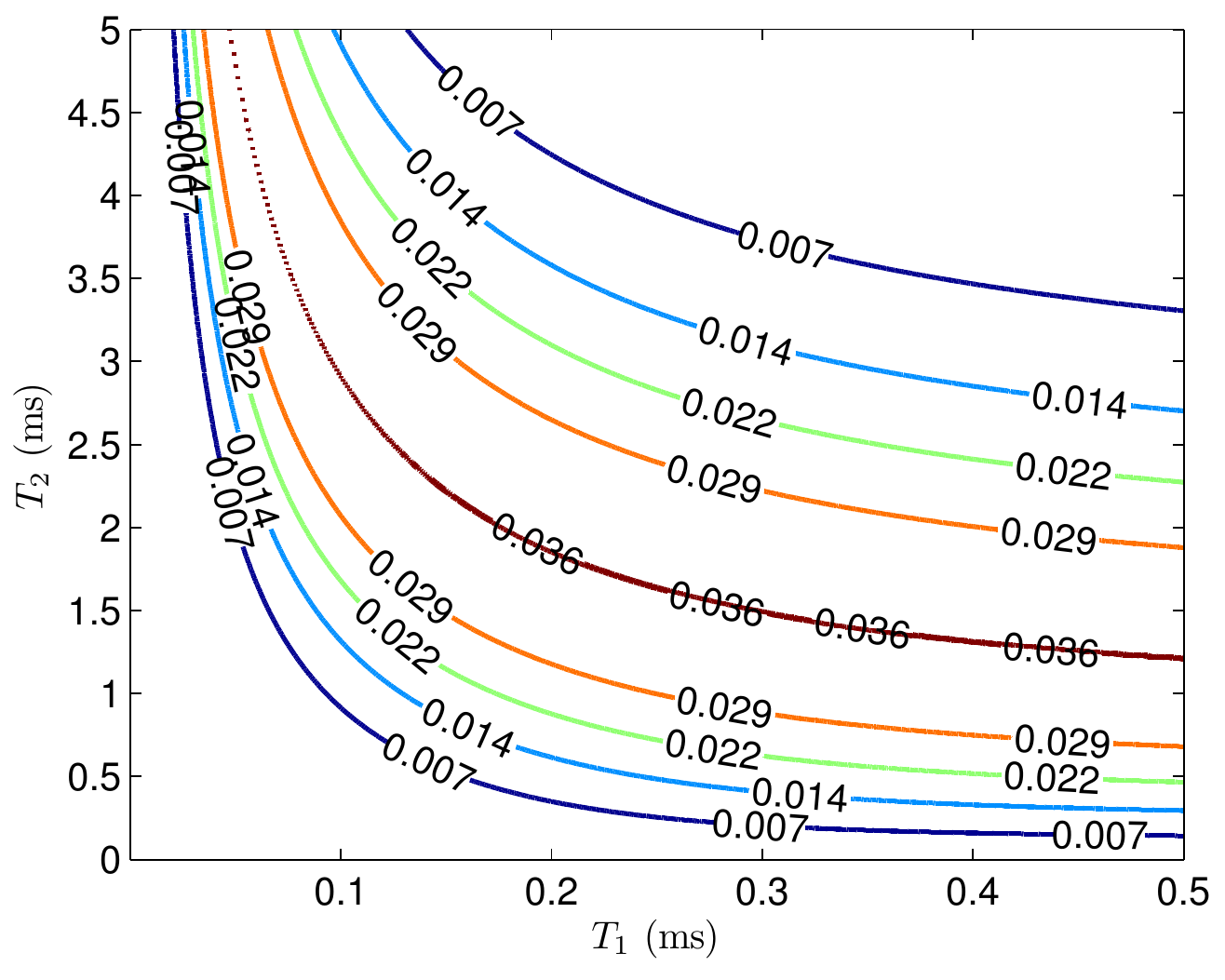}
	}
	\caption{The spatial throughput versus time durations of $T_1$ and $T_2$ without energy and information outage constraints. The energy threshold is $E_C=6~\mu\mathrm{J}$, and the targeted SINR level is $\gamma_R=5$ dB.}
	\label{fig_Thr_T}
\end{figure}
As we have indicated in Section \ref{sec_st_max}, the spatial throughput largely depends on the choice of the slot division ($T_1$ and $T_2$). We first show the  spatial throughput versus time durations of $T_1$ and $T_2$ in Fig. \ref{fig_Thr_T} without energy and information outage constraints.
It can be observed that for any given $T_1$ there does exist a $T_2$ that maximizes the spatial throughput and vice versa.
Since the optimal spatial throughput achieved by multiple $(T_1^\star,T_2^\star)$ pairs is the same from Fig. \ref{fig_Thr_T}, there exists a tradeoff between the optimal $T_1^\star$ and $T_2^\star$ as we have discussed in Section \ref{sec_st_max}.
From Fig. \ref{fig_Thr_T}, when $T_1$ becomes smaller in the regime of $T_1<0.1$ ms, the optimal $T_2^\star$ increases rapidly. The reason behind this is two-fold: On the one hand, since the time duration for energy harvesting is limited  the energy outage becomes the bottleneck of the spatial throughput, and a larger $T_2$ is beneficial to maintain low energy outage probability; On the other hand, since the density of the working tags reduces in the backscatter modulation phase, a longer time duration of $T_2$ is needed to transmit more information bits in the second phase.

\begin{figure}[t]
	\centering
	\subfloat[]{ 
		\includegraphics[width=3.5in]{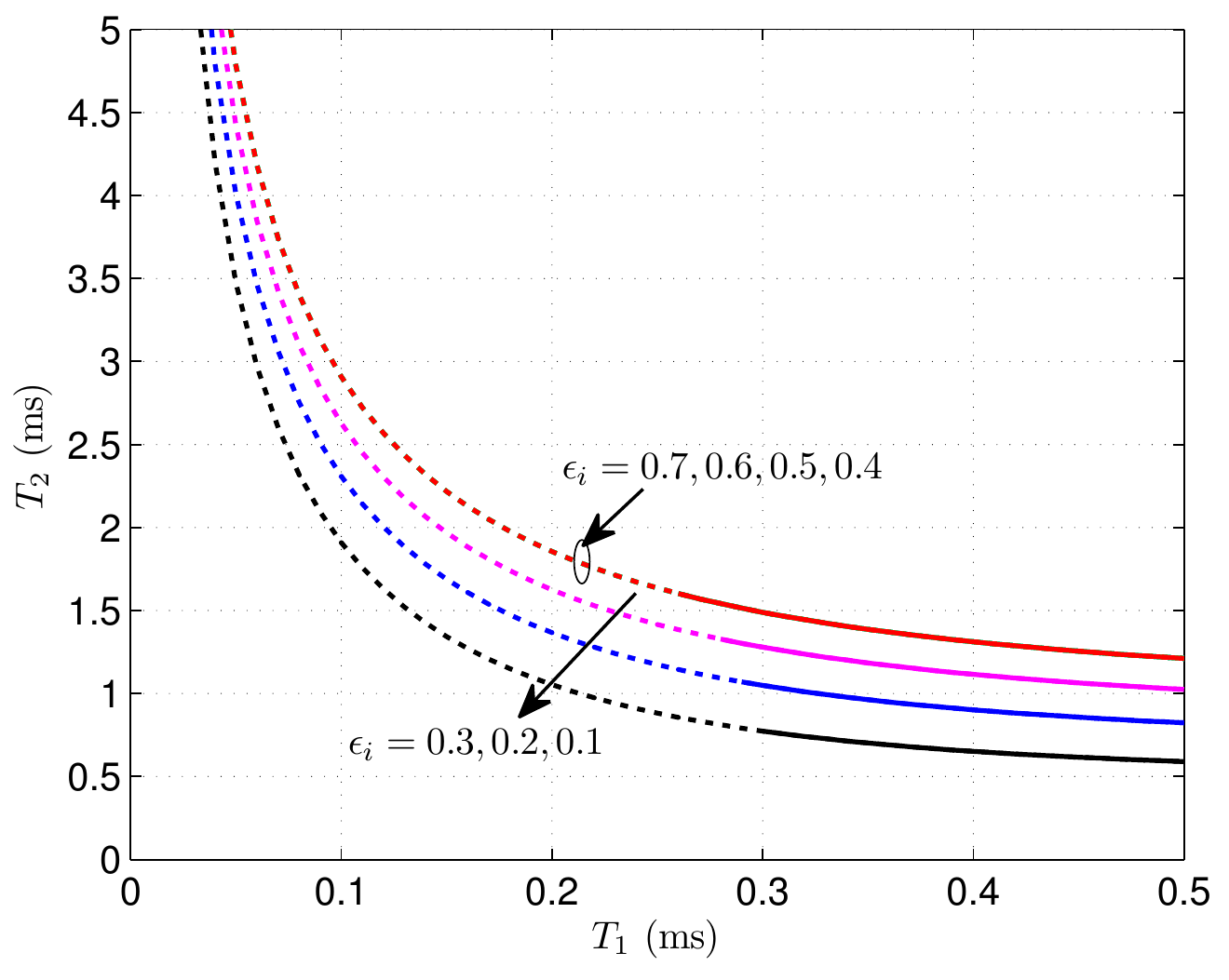}
	}
	\hfil
	\subfloat[]{ 
		\includegraphics[width=3.5in]{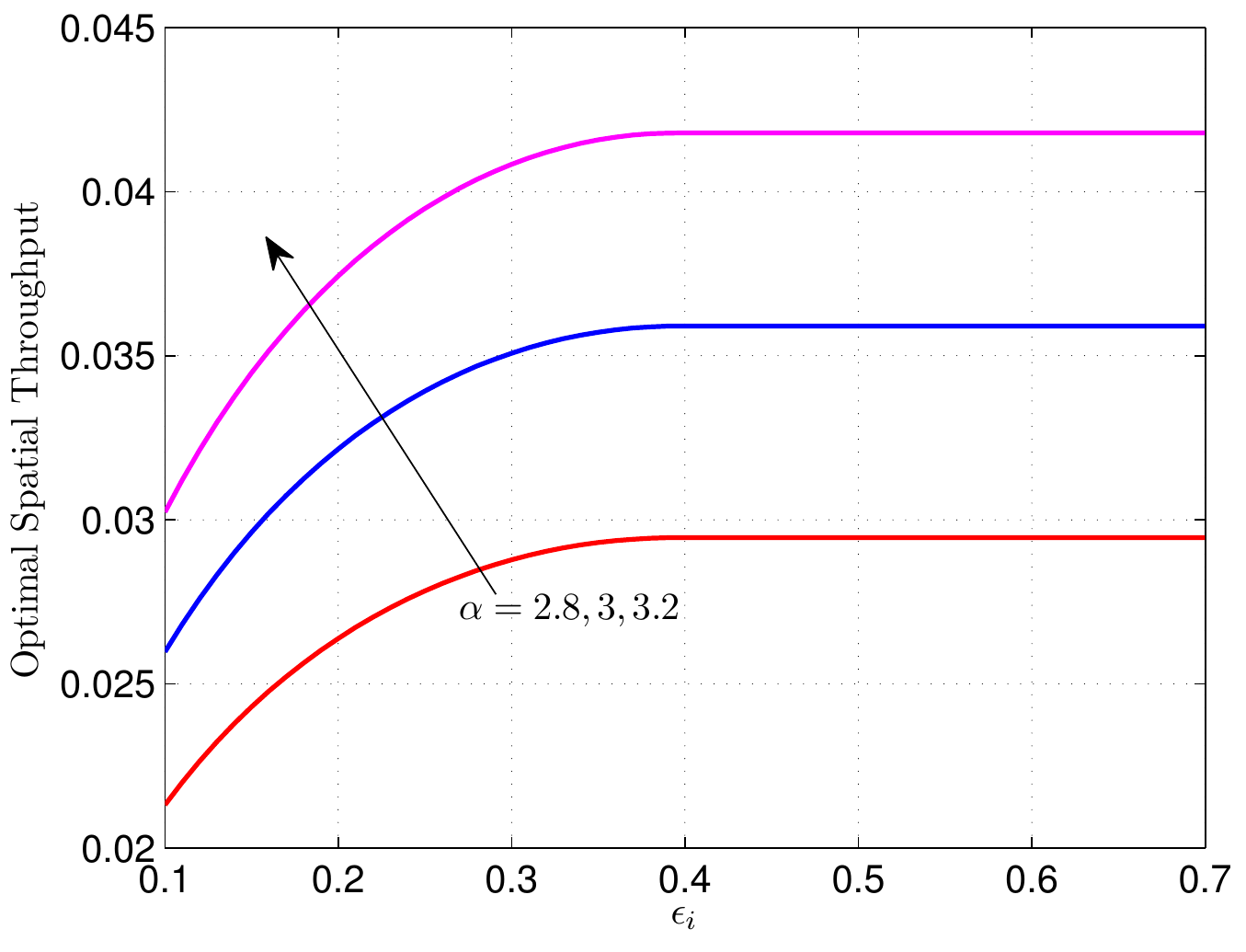}
	}
	\caption{(a) The optimal solution collections for $(T_1^\star,T_2^\star)$ and (b) the corresponding optimal spatial throughput under different constraints of information outage probability $\epsilon_i$'s. The solid lines in (a) stand for the situation where the constraint of energy outage probability is further taken into account over dashed lines. The constraint of energy outage probability is the same $\epsilon_e=0.4$. The energy threshold is $E_C=6~\mu\mathrm{J}$, and the targeted SINR level is $\gamma_R=5$ dB.}
	\label{fig_Opt_T}
\end{figure}
When the constraint of information outage probability is taken into consideration, the optimal solution collection for $(T_1^\star,T_2^\star)$ and the corresponding optimal spatial throughput are shown in Fig. \ref{fig_Opt_T}. 
The solid lines in Fig. \ref{fig_Opt_T}(a) stand for the situation where the constraint of energy outage probability is further taken into account over dashed lines.
From Fig. \ref{fig_Opt_T}(a), we can see that with the more and more stringent information outage constraint, the time durations for optimal $T_1^\star$ and $T_2^\star$ become smaller and the collection $\{(T_1^\star,T_2^\star)\}$ gets shrunk. This is because the information outage probability is an increasing function of $T_1$ and $T_2$ as we have pointed out in Section \ref{subsec_inf_out}.
In addition, when $\epsilon_i$ increases the optimal set for $T_2^\star$ becomes smaller so as to meet the same constraint for energy outage probability.
It is also straightforward to foresee that the optimal set for $T_2^\star$ will further get smaller when $\epsilon_e$ decreases according to \eqref{sol}.
Moreover, it is worth noting that from Fig. \ref{fig_Opt_T}(a) the optimal solution collections for $(T_1^\star,T_2^\star)$ are almost the same for $\epsilon_i=0.7,0.6,0.5,0.4$. This is because the information outage constraints are loose at this time and we have the same optimal $\chi^\star=x_M$.
Fig. \ref{fig_Opt_T}(b) shows that the optimal spatial throughput increases and finally remains at its maximal value with an increase in $\epsilon_i$. Additionally, a larger path-loss exponent $\alpha$ leads to a higher spatial throughput as observed from Fig. \ref{fig_Opt_T}(b).

\begin{figure}[t]
	\centering
	\includegraphics[width=3.5in]{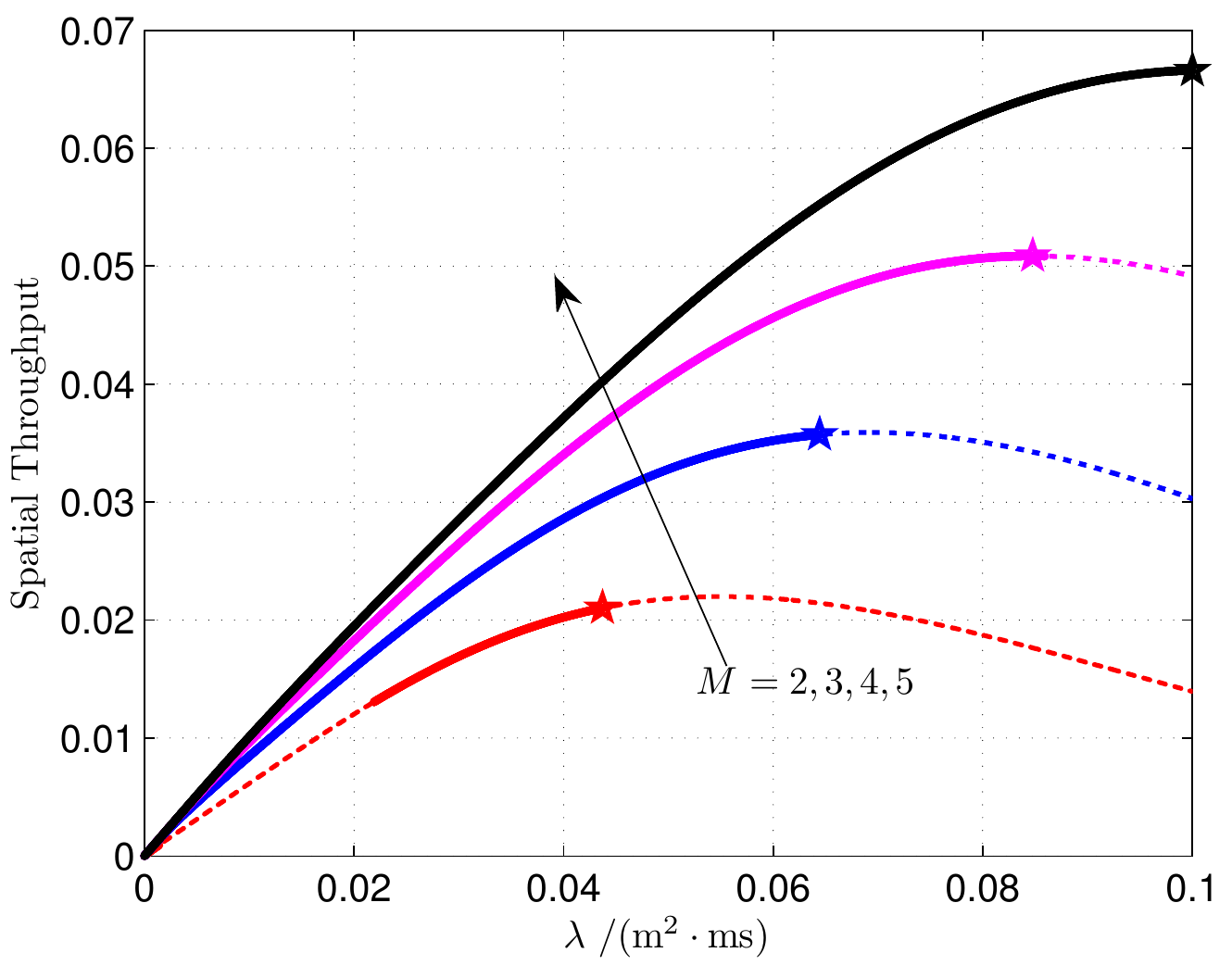}
	\caption{The spatial throughput versus the time-space density of the network $\lambda$ under different numbers of antennas equipped at the readers. The solid lines stand for the situation where the constraints of energy and information outage probabilities are further taken into account over dashed lines, and the star markers indicate the optimal $\lambda$ for spatial throughput maximization. The energy and information outage constraints are $\epsilon_e=\epsilon_i=0.35$. The energy threshold is $E_C=6~\mu\mathrm{J}$, and the targeted SINR level is $\gamma_R=5$ dB.}
	\label{fig_Thr_l}
\end{figure}
Fig. \ref{fig_Thr_l} shows the impact of the network time-space density on the performance of the spatial throughput. The solid lines in Fig. \ref{fig_Thr_l} stand for the situation where the constraints of energy and information outage probabilities are further taken into account over dashed lines, and the star markers further indicate the optimal network density.
It can be observed from Fig. \ref{fig_Thr_l} that there does exist an optimal $\lambda$ for spatial throughput maximization, since too large $\lambda$ induces a lot of interferences in the backscatter modulation phase while too small $\lambda$ limits the density of reader-tag pairs and may even incur the shortage of harvested energy in the energy harvest phase.
Moreover, the number of antennas equipped at the readers is critical to the network performance under the same energy and information outage constraints. Indeed, from Fig. \ref{fig_Thr_l} the feasible region of $\lambda$ becomes larger when $M$ increases from 2 to 5, since a larger $M$ is beneficial for both energy harvest and information reception. In addition, both the spatial throughput and the optimal network density increase as $M$ increases.

\section{Conclusion} \label{sec_conclusion}
In this paper, we have studied the network-wide performance of a multi-antenna B-WPC ad hoc network.
By exploiting a TS-PPP model, the behavior of the network with sporadic short packets has been well captured in a decentralized and asynchronous transmission way.
We have obtained the analytical energy and information outage probabilities in closed forms, respectively.
Furthermore, we have obtained the optimal transmission slot length and division by maximizing the network-wide spatial throughput. 
Finally, we have presented numerical results to verify our theoretical analysis.
It is shown that there exists the optimal tradeoff between the durations of the two transmission phases for spatial throughput maximization, and this tradeoff region gets shrunk when the outage constraints become more stringent.

\begin{appendices}
\section{Proof of Lemma \ref{Lemma_EDX}}\label{app_Lemma_EDX}
Since $\mb{h}_{x\hat{o}}$ is independent of $\mb{w}_{x}$, we have $\mb{h}_{x\hat{o}}^\T\mb{w}_{x}\triangleq \xi_x \sim \mathcal{CN}(0,1)$ and $\{\xi_x\}$ are independent and identically distributed (i.i.d.). 
Therefore, the mean of $Z_H$ can be found as
\begin{align}\label{EX1}
&\mathbb{E}\{E_H\} \notag\\
{=}&\eta \int_{0}^{T_1} \mathbb{E}\left\{|y_T(t)|^2\right\} dt \notag\\
\overset{\text{(a)}}{=}&\eta P_T \int_{0}^{T_1} \Bigg(G d_0^{-\alpha}+\mathbb{E}\Big\{\sum_{\substack{(x,t_x)\in\\ \Phi\backslash (o,0)}} e(t,t_x)  L(d_{x\hat{o}})^{-\alpha} \Big\}\Bigg)  dt \notag\\
\overset{\text{(b)}}{=}&\eta P_T  \int_{0}^{T_1} \Bigg( G d_0^{-\alpha} \notag\\
&\qquad+ 2\pi\lambda\int_{-\infty}^{\infty}\int_0^\infty e(t_1,t_2) L(r)^{-\alpha} rdrdt_2 \Bigg)  dt_1 \notag\\
{=}&\eta P_T T_1 \left(G d_0^{-\alpha}+ \pi \lambda T \frac{\alpha}{\alpha-2} r_o^{2-\alpha}\right),
\end{align}
where step (a) is obtained by noting that $\{\xi_x\}$ are i.i.d. $\mathcal{CN}(0,1)$ random variables, and step (b) is an application of the Campbell theorem\cite{Chiu2013}. 

Similarly, the second order moment of $E_H$ is given by
\begin{align}\label{EX2}
&\mathbb{E}\{E_H^2\} 
=(\eta P_T)^2\int_{0}^{T_1}\int_{0}^{T_1} \mathbb{E}\left\{q(t_1)q(t_2)\right\} dt_1 dt_2 ,
\end{align}
where
\begin{align} \label{h_t}
q(t)\triangleq 
& Gd_0^{-\alpha}+ 2\mathrm{Re}\Bigg\{\sqrt{Gd_0^{-\alpha}}\sum_{\substack{(x,t_x)\in\\ \Phi\backslash (o,0)}} e(t,t_x) \bar{\xi}_x L(d_{x\hat{o}})^{-\alpha/2}\Bigg\}\notag\\
&\quad+ \Big|\sum_{\substack{(x,t_x)\in\\ \Phi\backslash (o,0)}} e(t,t_x) \bar{\xi}_x L(d_{x\hat{o}})^{-\alpha/2}\Big|^2
\end{align}
with $\bar{\xi}_x=\xi_x e^{j(\varphi_x-\varphi_{\hat{o}})}$ following the $\mathcal{CN}(0,1)$ distribution.
By plunging \eqref{h_t} into \eqref{EX2} and leveraging the Campbell theorem\cite{Chiu2013}, we finally obtain $\mathbb{E}\{E_H^2\} $ after several algebraic manipulations as
\begin{align}\label{EX22}
&\mathbb{E}\{E_H^2\} \notag\\
&=(\eta P_T T_1)^2 \Bigg((Gd_0^{-\alpha} )^2+
\frac{2}{3} (5T_1+6T_2)\pi \lambda \frac{Gd_0^{-\alpha} \alpha}{\alpha-2} r_o^{2-\alpha} \notag\\
&~+\frac{2}{3} (2T_1+3T_2)\pi \lambda \frac{\alpha}{\alpha-1} r_o^{2(1-\alpha)}  \notag\\
&~+\left(\pi \lambda \frac{\alpha}{\alpha-2} r_o^{2-\alpha}\right)^2 \frac{1}{6} \left(9T_1^2+20T_1 T_2+12T_2^2\right)\Bigg).
\end{align}
Based on \eqref{EX1} and \eqref{EX22}, the variance of $Z_H$ can be found as
\begin{align}
&\mathbb{D}\{E_H\}\notag\\
&=\mathbb{E}\{E_H^2\}-\mathbb{E}^2\{E_H\} \notag\\
&=(\eta P_T T_1)^2 \Bigg(\frac{2}{3} (2T_1+3T_2)\pi \lambda \alpha r_o^{2-\alpha} \left(\frac{Gd_0^{-\alpha}}{\alpha-2}+\frac{r_o^{-\alpha}}{\alpha-1}\right)  \notag\\
&~+\left(\pi \lambda \frac{\alpha}{\alpha-2} r_o^{2-\alpha}\right)^2 \frac{1}{6} \left(3T_1^2+8T_1 T_2+6T_2^2\right)\Bigg).
\end{align}
The proof is completed.

\section{Proof of Theorem \ref{th_P_io}}\label{app_P_io}
The proof of \eqref{P_io_res} follows the similar line of that in \cite[Theorem 1]{Ali2010}, but here we consider a TS-PPP and further cope with time domain.
First, let us focus on the SINR performance in a given region of area $A$ over time $T_0$ denoted by the domain $(\mathcal{A},\mathcal{T}_0)\subseteq \mathbb{R}^2\times \mathbb{R}$.
We denote the number of the emerged reader-tag pairs in $(\mathcal{A},\mathcal{T}_0)$ as the random variable $N$ with the distribution given in \eqref{Pr_N}.
By applying the result in \cite{Gao1998}, the complementary cumulative distribution function (CCDF) of the $\mr{SINR}_R$ (or namely, the information connection probability) can be represented as\cite{Ali2010}
\begin{align}\label{F0}
&\bar{P}_{ic}(\lambda_t,\gamma_R)\notag\\
&\triangleq \Pr\{\mr{SINR}_R>\gamma_R\} \notag\\
&=\exp\left(-\sigma^2 \vartheta\right) \E_N\Bigg[ \sum_{i=0}^{M-1}\sum_{k=0}^{\min(i,N)} \binom{N}{k} \frac{(\sigma^2\vartheta)^{i-k}}{(i-k)!} \notag\\
&\qquad\qquad \times \E_X^k\left[\frac{\vartheta X}{1+\vartheta X}\right] \E_X^{N-k}\left[\frac{1}{1+\vartheta X}\right] \Bigg],
\end{align}
where $\sigma^2\triangleq \frac{N_0}{\beta \bar{y}_T^2 }$, $\vartheta\triangleq \gamma_R d_0^\alpha$, and $X\triangleq w(t_x) r_{\hat{x}o}^{-\alpha}$ is a random variable determined by $(\hat{x},t_x)\in (\mathcal{A},\mathcal{T}_0)$.
Recall from the property of PPP that once conditioned on $N$, $(\hat{x},t_x)$ is uniformly distributed in the domain $(\mathcal{A},\mathcal{T}_0)$\cite{Haenggi2012}.
Therefore, we further have
\begin{align}\label{E1}
\E_X\left[\frac{\vartheta X}{1+\vartheta X}\right]=\frac{1}{AT_0}\int_{(\mathcal{A},\mathcal{T}_0)} \frac{\vartheta X}{1+\vartheta X} d(\hat{x},t_x)
\end{align}
and
\begin{align}\label{E2}
\E_X\left[\frac{1}{1+\vartheta X}\right]=\frac{1}{AT_0}\int_{(\mathcal{A},\mathcal{T}_0)} \frac{1}{1+\vartheta X} d(\hat{x},t_x).
\end{align}
Substituting \eqref{E1}, \eqref{E2}, and the distribution of $N$ given in \eqref{Pr_N} into \eqref{F0} yields
\begin{align}
&\bar{P}_{ic}(\lambda_t,\gamma_R) \notag\\
&=\exp\left(-\sigma^2 \vartheta\right)\sum_{N=0}^{\infty} \sum_{i=0}^{M-1}\sum_{k=0}^{\min(i,N)} \binom{N}{k} \frac{(\sigma^2\vartheta)^{i-k}}{(i-k)!} \notag\\
&\quad \times \left(\lambda_t \int_{(\mathcal{A},\mathcal{T}_0)} \frac{\vartheta Xd(\hat{x},t_x)}{1+\vartheta X} \right)^k
\left(\lambda_t \int_{(\mathcal{A},\mathcal{T}_0)} \frac{d(\hat{x},t_x)}{1+\vartheta X} \right)^{N-k} \notag\\
&\quad \times 
\frac{\exp(-\lambda_t A T_0)}{N!} \notag\\
&=\exp\left(-\sigma^2 \vartheta\right) \sum_{i=0}^{M-1}\sum_{k=0}^{i}\sum_{N=k}^{\infty}  \frac{(\sigma^2\vartheta)^{i-k}}{k!(i-k)!} \notag\\
&\quad \times \left(\lambda_t \int_{(\mathcal{A},\mathcal{T}_0)} \frac{\vartheta Xd(\hat{x},t_x)}{1+\vartheta X} \right)^k
\frac{1}{(N-k)!}\notag\\
&\quad \times \left(\lambda_t \int_{(\mathcal{A},\mathcal{T}_0)} \frac{d(\hat{x},t_x)}{1+\vartheta X} \right)^{N-k}
\exp(-\lambda_t A T_0).
\end{align}
Using the fact that $ A T_0=\int_{(\mathcal{A},\mathcal{T}_0)}1d(\hat{x},t_x)$, we have
\begin{align}\label{F1}
&\bar{P}_{ic}(\lambda_t,\gamma_R) \notag\\
&=\exp\left(-\sigma^2 \vartheta\right) \sum_{i=0}^{M-1}\sum_{k=0}^{i} \frac{(\sigma^2\vartheta)^{i-k}(\lambda_t \varpi)^k}{k!(i-k)!}
\exp(-\lambda_t \varpi) \notag\\
&=\sum_{i=0}^{M-1} \frac{(\lambda_t \varpi+\sigma^2\vartheta)^i}{i!}
\exp(-\lambda_t \varpi -\sigma^2 \vartheta),
\end{align}
where 
\begin{align}\label{def_varpi}
\varpi \triangleq \int_{(\mathcal{A},\mathcal{T}_0)} \frac{\vartheta X}{1+\vartheta X} d(\hat{x},t_x).
\end{align}
Finally, by letting the formerly restricted domain $(\mathcal{A},\mathcal{T}_0)$ go to infinity, the integral $\varpi$ in \eqref{def_varpi} can be calculated as
\begin{align}\label{varpi}
\varpi
=& 2\pi \int_{-\infty}^{\infty} \int_{0}^{\infty} \frac{\vartheta w(t) r^{-\alpha}}{1+\vartheta w(t) r^{-\alpha}}r drdt \notag\\
\overset{\text{(a)}}{=}&\frac{2\pi }{\alpha}  \int_{0}^{\infty} \frac{x^{-\delta}}{1+x} dx \int_{-\infty}^{\infty} (\vartheta w(t))^\delta dt  \notag\\
\overset{\text{(b)}}{=}&\frac{2\pi }{\alpha}  \frac{\pi}{\sin(\delta\pi)} 2\vartheta^\delta \int_{0}^{T_2} \left(\frac{T_2-t}{T_2}\right)^\delta dt  \notag\\
{=}&\vartheta^\delta T_2 \Delta
\end{align}
where $\Delta\triangleq \frac{4\pi}{2+\alpha}\Gamma(\delta)\Gamma(1-\delta)=\frac{4\pi^2}{(2+\alpha)\sin(\delta\pi)}$, $\delta\triangleq \frac{2}{\alpha}$, step (a) follows from the variable change $x= \vartheta w(t) r^{-\alpha}$, and equality (b) is obtained by the use of \cite[(3.222.2)]{Gradshteyn2007} and \eqref{w_tx}.
Plunging \eqref{varpi} into \eqref{F1} yields the result shown in Theorem \ref{th_P_io}.
The proof is completed.

\end{appendices}


\end{document}